\documentclass[twocolumn,showpacs,preprintnumbers,amsmath,amssymb,floatfix,aps,nofootinbib]{revtex4-1}
\usepackage{color}   
\usepackage{graphicx}
\usepackage{dcolumn}
\usepackage{bm}
\usepackage{slashed}

\begin{document}

\def\bb    #1{\hbox{\boldmath${#1}$}}

\title{Tsallis Fits to $p_T$ Spectra and Multiple Hard Scattering in
  pp Collisions at LHC} 
\author{Cheuk-Yin Wong}
\email{wongc@ornl.gov}

\affiliation{Physics Division, Oak Ridge National Laboratory, 
Oak Ridge, Tennessee 37831, USA}
\author{Grzegorz Wilk}
\email{wilk@fuw.edu.pl}

\affiliation{ National Centre for Nuclear Research, Warsaw 00-681, Poland}

\date{\today}

\begin{abstract}
Phenomenological Tsallis fits to the CMS, ATLAS, and ALICE transverse
momentum spectra of hadrons for $pp$ collisions at LHC were recently
found to extend over a large range of the transverse
momentum. We investigate whether the few degrees of freedom in the
Tsallis parametrization may arise from the relativistic
parton-parton hard-scattering and related processes.  The effects of
the multiple hard-scattering and parton showering processes on the power law are
discussed.  We find empirically that whereas the transverse spectra of both
hadrons and jets exhibit power-law behavior of $1/p_T^n$ at high
$p_T$, the power indices $n$ for hadrons are systematically greater
than those for jets, for which $n$$\sim$4-5.

\end{abstract}


\maketitle

\vspace{1cm}

\section{Introduction}

The transverse momentum distributions of produced particles in hadron
and nuclear collisions provide useful information on the dynamics of
the colliding system.  The low-$p_T$ part of the spectra falls within
the realm of soft nonperturbative QCD physics and may involve the
parton wave functions in a flux tube \cite{Gat92}, the
thermodynamics\footnote{ The usual (extensive) thermodynamics with the
  Boltzmann-Gibbs distribution have been described in
  \cite{H,HR,Kra09} and applied extensively for multiparticle production
  in \cite{Hwa03}.  Its nonextensive generalization with the Tsallis
  distribution with a new nonextensivity parameter $q$ has been given
  in \cite{Tsallis}. The nonextensive statistical approach has been
  very successful in describing many different physical systems,
  including multiparticle production processes at lower energies.  See
  Refs. \cite{Wil12b,WW,Wil12a,Wibig,Urm12,ClW,View2} for a
  summary of earlier attempts to use Tsallis fits and detailed
  explanations of the possible meaning of the $q$ parameter. } and the
recombination of partons
\cite{H,HR,Kra09,Hwa03,Tsallis,Wil12b,WW,Wil12a,Wibig,Urm12,ClW,H,View2},
or the fragmentation of a QCD string \cite{And83}.  On the other hand,
the high-$p_T$ part is usually considered to arise from a perturbative
QCD hard-scattering between a parton of one hadron and a parton of the
other hadron \cite{Bla74,Bla75,Sch77,Bla77,Won94,Sjo87,Arl10,Bro12}. The
borderline between the soft-$p_T$ nonperturbative region and the
high-$p_T$ perturbative region is not well determined.  A very
different scheme to partition the $p_T$ spectrum into soft and hard
components has also been suggested \cite{Ada06,Tra08} and will be
discussed at the end of this paper.

In recent RHIC and LHC experiments, the transverse momentum spectra of
charged hadrons for $pp$ and nucleus-nucleus collisions have been
measured at very high energies
\cite{STAR,PHENIX,ATLAS,CMS11,CMS12a,ALICE}.  These spectra are often
described by the Tsallis distribution \cite{Tsallis}
\begin{equation}
h_q\left( p_T\right) = C_q \left[ 1 -
(1-q)\frac{p_T}{T}\right]^{\frac{1}{1-q}},\label{eq:Tsallis}
\end{equation}
with a normalization constant $C_q$, a ``temperature" $T$, and a
dimensionless nonextensivity parameter $q$ (with $q > 1$).  The
Tsallis distribution can be regarded as a nonextensive generalization
of the usual exponential (Boltzmann-Gibbs) distribution, and converges
to it when the parameter $q$ tends to unity,
\begin{equation}
 h\left( p_T\right)
 \stackrel{q
\rightarrow 1}{\Longrightarrow} C_1\exp
\left(-\frac{p_T}{T}\right).
\end{equation}
It has been very successful in describing very different physical
systems in terms of a statistical approach, including multiparticle
production processes at lower energies.  
\cite{Wil12b,WW,Wil12a,Wibig,Urm12,ClW,View2}. 

On the other hand, long time ago Hagedorn proposed the {\it QCD
  inspired} empirical formula to describe experimental hadron
production data as a function of $p_T$ over a wide range \cite{H}:
\begin{eqnarray}
  E\frac{d^3\sigma}{d^3p}& = &C\left( 1 + \frac{p_T}{p_0}\right)^{-n}
\nonumber\\
 & \longrightarrow&
  \left\{
 \begin{array}{l}
  \exp\left(-\frac{n p_T}{p_0}\right)\quad \, \, \, {\rm for}\ p_T \to 0, \smallskip\\
  \left(\frac{p_0}{p_T}\right)^{n}\qquad \qquad{\rm for}\ p_T \to \infty,
 \end{array}
 \right .
 \label{eq:H2}
\end{eqnarray}
where $C$, $p_0$, and $n$ are fitting parameters. This becomes a
purely exponential function for small $p_T$ and a purely power-law
function for large $p_T$ values\footnote{Actually the QCD formula was
  inspired by related work in \cite{Bla74,Bla75,Sch77,Bla77} and proposed
  earlier in \cite{CM,UA1}}. It coincides with Eq.  (\ref{eq:Tsallis})
for
\begin{equation}
n = \frac{1}{q - 1}\quad {\rm and}\quad p_0 = \frac{T}{q -
1}.\label{eq:coincides}
\end{equation}
Usually both formulas are treated as equivalent from the point of view
of phenomenological fits and are often used interchangeably
\cite{STAR,PHENIX,ATLAS,CMS11,CMS12a,ALICE}. It
is worth stressing that both Eq. (\ref{eq:Tsallis}) and Eq.
(\ref{eq:H2}) describe data {\it in the whole region of transverse
momenta}, not only for large $p_T$.

For phenomenological as well as theoretical interests, it is expected
that as the low-$p_T$ region and the high-$p_T$ region arise from
different mechanisms, there can be a change of the systematics for the
description of the low-$p_T$ nonperturbative QCD region and the
high-$p_T$ perturbative QCD region.  It is therefore useful to explore
where the Tsallis fit begins to fail at higher and higher $p_T$ in the
recent high-$p_T$ data of CMS \cite{CMS11,CMS12a}, ATLAS \cite{ATLAS},
and ALICE Collaborations \cite{ALICE} for $pp$ collisions at the LHC.
An excellent fit to the $p_T$ hadron spectra was earlier obtained there
with the Tsallis and/or Hagedorn distributions for $p_T$ from 0.5 GeV
up to 6 GeV, in $pp$ collisions at $\sqrt{s}=$7 TeV \cite{CMS11}.  It
was however a surprise to find that the phenomenological Tsallis fits
to the CMS and ATLAS charged particle transverse spectra extends from
$p_T$=0.5 to 181 GeV/c in $pp$ collisions at $\sqrt{s}=$7 TeV, and
from $p_T$=0.5 to 31 GeV/c at $\sqrt{s}=$0.9 TeV \cite{Won12}.  The
simplicity of the Tsallis parametrization with only three parameters
and the large range of the fitting transverse momentum raise questions
on the physical meaning of the degrees of freedom that enter into the
high-$p_T$ distribution.

As the magnitude of the transverse momenta in these high-$p_T$ data
are much greater than the mean transverse momentum of the
distribution, concepts such as statistical mechanics that depend on
thermodynamical equilibrium or quasiequilibrium may be subject to
question.  The asymmetry between the transverse and the longitudinal
degrees of freedom also poses additional difficulties in a statistical
explanation of the full three-dimensional momentum
distribution\footnote{ However, it should be remembered that
  statistical approach is not the only known source of Tsallis
  distribution in Eq. (\ref{eq:Tsallis}). There are numerous dynamical
  mechanisms leading to it, see \cite{WW,qNet,qCas}.}.

To describe the transverse momentum distribution in the high $p_T$
region, a more natural description would be to employ the relativistic
hard-scattering model in perturbative QCD.  We wish to investigate
whether the few degrees of freedom in the transverse momentum Tsallis
distribution may arise from the basic parton-parton scattering and the
accompanying multiple collision and showering processes.

The relativistic hard-scattering model has been used previously to
examine inclusive particle production in hadron-hadron collisions
\cite{Bla74,Bla75,Sch77,Bla77,Won94,Sjo87, Arl10,Bro12}.  It was found
earlier on that the observed experimental hadron transverse
differential cross section appears to differ from what one expects
from naive point parton collisions.  In the basic quark model, the
high-$p_T$ differential cross section in an $ab \to cd$ exclusive
process can be inferred from the counting rule of Brodsky, Farrar,
Matveev $et~al.$ \cite{Bro73,Mat73}, which states that the invariant
cross section for the exclusive process at high-$p_T$ behaves as the
power law, with power index $n$,
\begin{eqnarray}
E_c \frac{d\sigma (ab \to cd)}{dc^3}
\propto \frac{1}{c_T^n},
\end{eqnarray}
where $n=2\times \{({\rm number ~of~active~participants}) -2\}$.  The
counting of the number of active participants includes constituents in
the initial $ab$ and the final $cd$ states.  (For a pedagogical
discussion of the counting rule, see \cite{Won94}.)  The counting rule
of Brodsky, Farrar, Matveev $et~al.$ \cite{Bro73,Mat73} has been found
to give a power index $n$ that agrees reasonably with experimental data
for exclusive $ab \to cd$ processes \cite{Whi94}.  If one assumes that
the dominant basic high-$p_T$ parton-parton hard-scattering process in
a $pp$ collision comes from $qq \to qq$ (or other $2\to 2$ processes),
then the counting rule gives a transverse momentum dependence of
$d\sigma/dt\sim 1/p_T^n$ with $n=4$.  However, the observed
experimental power index $n$ of the hadron transverse spectrum is
about 7 (even at the highest LHC energy and for very large transverse
momenta measured \cite{Won12}).  If one assumes that the basic process
is $q+$meson $\to$ $q$+meson, then the counting rule gives $n=8$ which
is close to the observed value.  Blankenbecler, Brodsky and Gunion
therefore proposed that the power index of $n\sim 8$ may be related to
the scattering of a parton with a meson \cite{Bla75,Sch77,Bla77}.  For
$pp$ collisions at the LHC, a modified proposal with the direct meson
production in the basic reaction $g + q \to $meson+$q$ has been
suggested recently, involving $5$ active participants and $n=6$ for
the power index \cite{Arl10,Bro12}.

We will however not work with mesons as elementary participant
constituents as in \cite{Bla75,Sch77,Bla77,Arl10,Bro12} but will work
within the conventional parton model of quarks and gluons.  The
collision of hadrons (or nuclei) consists of the collisions of partons
either in parallel or in series.  For example, in the PYTHIA
Monte-Carlo program, the multiple hard scattering of partons in
parallel is an important ingredient and the number of hard-scattering
interactions per inelastic event may be greater than unity
\cite{Sjo87}.  The other process of multiple scattering of partons in
series has been examined in great detail previously
\cite{Kas87,Cal90,Cal90a,Cal91,Cal94,Cal94a,Cal01,Acc01,Gyu01,Cor10}.
Remarkably, a simple picture emerges from these studies to indicate
that as a result of the
multiple scattering, the sum of the multiple collision series in a
minimum-biased sampling 
at high $p_T$
is dominated by the differential cross section
for the single parton-parton scattering.  As a result of shadowing
cancellations, the high-$p_T$ scattering appears as though it arises
from a single scattering with a $1/p_T^4$ distribution, plus
logarithmic residue terms.  This remarkable result was shown in
\cite{Acc01}, using an auxiliary generating functional.  We would like
to follow and extend the multiple hard-scattering results of
\cite{Acc01}, in order to obtain an explicit form of the multiple scattering power law and logarithmic residue terms, the dependence on the number of partons, the dependence
on the number of scatterers, and the the dependence on the centrality
of the collision.  These new results may find applications in the
multiple hard-scattering processes in hadron-hadron as well as
nucleus-nucleus collisions.

Whereas the theoretical analyses of
\cite{Kas87,Cal90,Cal90a,Cal91,Cal94,Cal94a,Cal01,Acc01,Gyu01,Cor10}
indicate that the multiple scattering process involving partons will
not significantly modify the $1/p_T^4$ distribution of the high-$p_T$
transverse differential cross section with $n$=4, the PYTHIA program
with properly tuned sets of parameters in a relativistic
hard-scattering model, with the additional processes of parton
showering and radiations, can describe quite well the transverse
momentum distribution of produced hadrons in $pp$ collisions at LHC
energies \cite{CMS11} with $n$$\sim$7 \cite{Won12}.  What is the
origin of such a difference in the power indices $n$?  Could the
additional process of parton showering and hadronization affect the
power index $n$?

The possibility that parton showering and hadronization may influence
the power index $n$ is revealed by the measurements of the
transverse differential cross section of hadron and photon jets for $p\bar p$ collisions at Fermilab  by
the CDF and D0 collaborations \cite{Abe93,Abb01,Aco02,Abb00,Aba01}.  In these measurements, the power indices $n$ are found
to be close to $n$=4--5 (see Fig. 2 of \cite{Arl10}), as predicted from
perturbative QCD. A hadron jet in these measurements corresponds to a
collection of hadrons in calorimeter cells contained within a cone
of opening angle $R$, and it represents a parton after a parton-parton
collision but before its fragmentation and hadronization.  Its
transverse momentum differential cross section retains the main
features of the power law of $1/p_T^4$ of the basic parton-parton
hard scattering.  Thus, the difference between the power index of
$n$$\sim$4-5 from the jet transverse differential cross section and
$n$$\sim$7 from the hadron spectra is likely related to the subsequent
showering and hadronization of the parton jets to hadron fragments of
lower transverse momenta.  We would like to examine here how the
additional process of parton fragmentation and parton showering may
influence the power index of the transverse differential cross section.

This paper is organized as follows.  In Sec. II, we review the
relativistic hard-scattering model to express the scattering cross
section for high-$p_T$ processes in terms of the basic parton-parton
differential cross sections.  An approximate analytical expression is
obtained by carrying out the hard-scattering integral analytically.
In Sec. III, we study the effects of multiple hard-scattering of
partons on the differential cross sections.  In Sec. IV, we include
the effects of the additional dependence of the parton thickness
function $T(b)$ on the parton differential cross sections.  In Sec.
V, we analyze the experimental results of jet transverse differential
cross sections with the relativistic hard-scattering model and find the
approximate validity of the RHS model for jet production.  In Sec.
VI, we examine the effect of fragmentation on the hadron differential
cross section.  In Sec. VII, we study the effects of showering and
its effects on the power index.  In Sec. VIII, we fit the
experimental CMS, ATLAS, and ALICE data to the hard-scattering model
and extract the power index from data.  In Sec. IX, we present our
discussions and conclusions.

\section{Relativistic hard scattering model}

We review some of the earlier results in the relativistic hard
scattering model \cite{Bla74,Bla75,Sch77,Bla77,Won94,Won98}.  We
consider the process of $A+B \to c+X$ with the production of parton
$c$ around $\eta\sim 0$ in the center-of-mass frame of the $A$-$B$ system.  We shall later consider the fragmentation of the parton $c$ in
Sec. V and the showering process in Sec. VI.  The differential cross section
for this process is given in the parton model by

\begin{eqnarray}
E_c&&\frac{d^3\sigma( AB \to c X) }{dc^3}
=\sum_{ab} \int dx_a d{\bb a}_T dx_b d{\bb b}_T{~~~~}
\nonumber\\
&&~\times G_{a/A}(x_a,{\bb a}_T) G_{b/B} (x_b,{\bb b}_T) 
E_c \frac{d^3\sigma( ab \to c X') }{dc^3}.
\end{eqnarray}
We consider the basic process to be the lowest-order elastic
parton-parton collisions in which the parton-parton invariant cross
section is related to $d\sigma/dt$ by
\begin{eqnarray}
E_c \frac{d^3\sigma( ab \to c X') }{dc^3} &=&
 \frac{\hat s}{\pi}\frac{
  d\sigma( ab \to cX') } {dt} \delta (\hat s +\hat t +\hat u ),
\label{6}
\end{eqnarray}
where we have neglected the rest masses  and we have introduced
\begin{eqnarray}
\hat s = (a+b)^2,\nonumber\\
\hat t = (a-b)^2,\nonumber\\
\hat u = (b-c)^2.\nonumber
\end{eqnarray}
We write out the momenta in the infinite momentum frame, with
$\sqrt{s}$ the center-of-mass energy of $A$-$B$ system,
\begin{eqnarray}
a&=&(x_a \frac{\sqrt{s}}{2} + \frac{a_T^2}{2x_a \sqrt{s}}, ~{\bb a}_T, ~x_a \frac{\sqrt{s}}{2} - \frac{a_T^2}{2x_a \sqrt{s}}),
\nonumber\\
b&=&(x_b \frac{\sqrt{s}}{2} + \frac{b_T^2}{2x_b \sqrt{s}}, ~{\bb b}_T, -x_b \frac{\sqrt{s}}{2} + \frac{b_T^2}{2x_b \sqrt{s}}),
\nonumber\\
c&=&(x_c \frac{\sqrt{s}}{2} + \frac{c_T^2}{2x_c \sqrt{s}}, ~{\bb c}_T, ~x_c \frac{\sqrt{s}}{2} - \frac{c_T^2}{2x_c \sqrt{s}}) .\nonumber
\end{eqnarray}
The light-cone variable  $x_c$ of the produced parton   $c$ is
\begin{eqnarray}
x_c=\frac{c_0+c_z}{\sqrt{s}}.
\end{eqnarray}
The Mandelstam variables are
\begin{eqnarray}
\hat s&=&(a+b)^2
=x_a x_b {s}+\frac{a_T^2 b_T^2}{x_a x_b s}
-2 {\bb a}_T \cdot {\bb b}_T,\nonumber\\
\hat t&=&(a-c)^2= 
-\frac{ x_a c_T^2}{x_c }
-\frac{ x_c a_T^2}{x_a }
+ 2 {\bb a}_T \cdot {\bb c}_T,\nonumber\\
\hat u&=&(b-c)^2=
-{ x_b x_c s }
-\frac{  b_T^2 c_T^2 }{x_b x_c s }
+ 2 {\bb b}_T \cdot {\bb c}_T.\nonumber
\end{eqnarray}
The relation of $\hat s +\hat t + \hat u =0$ gives
\begin{eqnarray}
x_a x_b {s}&+&\frac{a_T^2 b_T^2}{x_a x_b s}
 -\frac{ x_a c_T^2}{x_c }
-\frac{ x_c a_T^2}{x_a }
-{ x_b x_c s }
-\frac{  b_T^2 c_T^2 }{x_b x_c s }
\nonumber\\
&=& -a_T^2-b_T^2-c_T^2 
+( \bb c_T - {\bb a}_T + {\bb b}_T )^2.
\end{eqnarray}
Because the intrinsic $a_T$ and $b_T$ are small compared with the
magnitudes of $a_z$, $b_z$, and $c_T$, we can therefore neglect terms
with $ a_T$ and $ b_T$ in the evaluation of $\hat s$, $\hat t$, and
$\hat u$.  We get
\begin{eqnarray}
\hat s = x_a x_b s,~~~ \hat t = -\frac{x_a c_T^2}{x_c},~~~\hat u=-x_b x_c s.
\end{eqnarray}
The constraint of $\hat s +\hat t + \hat u=0$ gives
\begin{eqnarray}
&&x_a(x_b)=x_c + \frac{c_T^2}{ (x_b 
 -\frac{  c_T^2}{x_c s})s}.
\end{eqnarray}
In the special case of particle $c$ coming out at
$\theta_c= 90^{\rm o}$ 
in the center-of-mass frame of the $A$-$B$ system,
\begin{eqnarray}
x_c&=&\frac{c_T}{\sqrt{s}},
~~~~x_a(x_b)=x_c +\frac{x_c^2}{x_b-x_c},
\end{eqnarray}
and 
\begin{eqnarray}
x_a=x_b=2x_c.
\end{eqnarray} 
The constraint in Eq.\ (\ref{6}) can be written as a constraint in
$x_a$,
\begin{eqnarray}
\delta(\hat s +\hat t + \hat u)=
\frac{\delta(x_a-x_a(x_b))}
{|\frac{\partial (\hat s +\hat t + \hat u)}{\partial x_a}|}.
\end{eqnarray}
On the other hand,
\begin{eqnarray}
\frac{\partial (\hat s +\hat t + \hat u)}{\partial x_a}
=s(x_b-\frac{c_T^2}{x_c s}).
\end{eqnarray}
We have therefore
\begin{eqnarray}
E_c \frac{d^3\sigma( ab \to c X') }{dc^3}
&=&
\frac{ d\sigma( ab \to cX) }
{dt}
\nonumber\\
& &\times
\frac{x_a x_b  \delta (x_a - x_a(x_b))}{\pi (x_b-c_T^2/x_c s)},
\end{eqnarray}
and 
\begin{eqnarray}
E_c \frac{d^3\sigma( AB \to c X) }{dc^3}
&=&\sum_{ab} \int d{\bb a}_T  d{\bb b}_T dx_b dx_a
\nonumber\\
&\times&
G_{a/A}(x_a,{\bb a}_T) G_{b/B} (x_b,{\bb b}_T)
\label{eq25}\\
&\times&
\frac{x_a x_b \delta (x_a - x_a(x_b))}{\pi (x_b-c_T^2/x_c s)} 
\frac{d\sigma( ab \to cX')}{dt}.\nonumber
\end{eqnarray}
We consider an approximate structure function of the form
\begin{eqnarray}
G_{a/A}(x_a,{\bb a}_T)=\frac{A_a}{x_a}(1-x_a)^{g_a}D_a(\bb a_T),\nonumber\\
G_{b/B}(x_b,{\bb b}_T)=\frac{A_b}{x_b}(1-x_b)^{g_b}D_b(\bb b_T).\nonumber
\end{eqnarray}
The integral in Eq.\ (\ref{eq25}) becomes
\begin{eqnarray}
&&E_c \frac{d^3\sigma( AB \to c X) }{dc^3}
=\sum_{ab} A_a A_b \int d{\bb a}_T  d{\bb b}_T  D_a(\bb a_T) D_b(\bb b_T)
\nonumber\\
&&\times  dx_b dx_a (1\!-\!x_a)^{g_a}(1\!-\!x_b)^{g_b}
\frac{ \delta (x_a - x_a(x_b))}{\pi (x_b-c_T^2/x_c s)}
\frac{d\sigma(ab \to cX')}{dt}.\nonumber
\end{eqnarray}
We integrate over $x_a$, and we get
\begin{eqnarray}
E_C \frac{d^3\sigma( AB \to c X) }{dc^3}
&&=\sum_{ab}{ A_a A_b}\int d{\bb a}_T  d{\bb b}_T  
D_a(\bb a_T) D_b(\bb b_T)      
\nonumber\\
&&\times  dx_b 
\frac{(1 \! -\! x_a)^{g_a}(1\! -\! x_b)^{g_b}}{\pi (x_b-c_T^2/x_c s)}
\frac{d\sigma(ab\! \to\! cX')}{dt}.\nonumber
\end{eqnarray}
As the transverse momentum we are considering is considerably larger
than the intrinsic $p_T$ \cite{Won98}, we can take the intrinsic
momentum distribution to be quite narrow so that the integration of
$\int d\bb a D_a(\bb a_T) = \int d\bb b D_b(\bb b_T)=1$ and we obtain
\begin{eqnarray}
&&E_C \frac{d^3\sigma( AB\!\! \to\!\! c X) }{dc^3}
=\sum_{ab}\!{ A_a A_b}\!\int\!  dx_b 
\frac{(1-x_a)^{g_a}(1-x_b)^{g_b}}{\pi (x_b-\tau_c^2)}\nonumber\\
&&
\hspace*{3.2cm}
\times \frac{d\sigma(ab \to cX')}{dt},
\end{eqnarray}
where we have introduced 
\begin{eqnarray}
\tau_c^2=\frac{c_T^2}{s}.
\end{eqnarray}  We use saddle point
integration method \cite{Won98} and get
\begin{eqnarray}
E_C \frac{d^3\sigma( AB \to c X) }{dc^3}
&&=\sum_{ab}{ A_a A_b} \int  dx_b 
\frac{e^{f(x_b)}}{\pi (x_b-\tau_c^2/x_c )}\nonumber\\
&&\times \frac{d\sigma(ab \to cX')}{dt},
\end{eqnarray}
with 
\begin{eqnarray}
f(x_b)=g_a \ln (1-x_a) + g_b \ln (1-x_b).\nonumber
\end{eqnarray}
Consider $g_a=g_b=g$ and expand $f(x_b)$ as a function of $x_b$ about
the minimum located at
\begin{eqnarray}
 x_{b0}  =\frac{\tau_c^2}{x_c}+  \tau_c  \sqrt{
\frac{1-\tau_c^2/x_c }{1-x_c}}.
\label{44}
\end{eqnarray}
The quantity $x_a$ at this minimum  is
\begin{eqnarray}
x_{a0}&
=&x_c  +  \tau_c\sqrt{\frac{1-x_c}{1-\tau_c^2/x_c}}.
\label{44a}
\end{eqnarray}
From the second derivative of $f(x_b)$ with respect to $x_b$, we
obtain
\begin{eqnarray}
E_c && \frac{d^3\sigma( AB \to c X) }{dc^3}
\sim 
\sum_{ab} \frac{A_a A_b}{\sqrt{\pi g_a}}
(1-x_{a0})^{g_a}(1-x_{b0})^{g_a}
\nonumber \\
&& \times \frac{1}{\sqrt{\tau_c}}
 \left \{
\frac{1-x_c}{1-\tau_c^2/x_c }\right \}^{1/4}
\sqrt{\frac{(1-x_{b0})^2}
{ [(1- (x_{b0}+ \tau_c^2/x_c)/2]  }
 } 
\nonumber\\
&&\times 
\frac{d\sigma(ab \to cX')}{dt}\biggr |_{x_{a0},x_{b0}}.
\end{eqnarray}
In the neighborhood of $\theta_c \sim 90^{\rm o}$ in the $A$-$B$
center-of-mass system, the ratios in the square-root factor and the
factor involving the power $1/4$ are approximately equal to 1.  Thus,
the analytical integration of the hard-scattering integral leads to
the following invariant differential cross section in an analytical
form,
 \begin{eqnarray}
E_c && \frac{d^3\sigma( AB \to c X) }{dc^3}
\sim
\sum_{ab} \frac{A_a A_b}{\sqrt{\pi g_a}}
(1-x_{a0})^{g_a}(1-x_{b0})^{g_a} \nonumber\\
&& \times 
\frac{1}{\sqrt{\tau_c}}
\frac{d\sigma(c_T;ab \to cX')}{dt}\biggr |_{x_{a0},x_{b0}}.
\label{rhs}
\end{eqnarray}

As an example, we can consider the basic $ab \to cX'$ process to be
$gg \to gg$.  The cross section as given by Gastman and Wu
\cite{Gas90} (page 403) is
\begin{eqnarray}
\frac{d\sigma (gg\to gg)}{dt}
&=&\frac{9\pi \alpha_{\hat s}^2}{8} \frac{({\hat s}^4+{\hat t}^4+{\hat u}^4)(({\hat s}^2+{\hat t}^2+{\hat u}^2)}{{\hat s}^4 {\hat t}^2 {\hat u}^2}.\nonumber
\end{eqnarray}
At $\theta\sim 90^{\rm o}$, we have
\begin{eqnarray}
\frac{d\sigma(gg\to gg) }{dt}
&=& \frac{9\pi \alpha_s^2}{16c_T^4}
\left [1 +\left  ( \frac{c_T^2}{x_c x_b s}\right )^2 
+ \left  ( \frac{x_c}{x_a} \right )^2  \right ]^3
\nonumber\\
&\sim& 
\frac{9\pi \alpha_s^2}{16c_T^4}
\left [\frac{3}{2} \right ]^3.
\label{36}
\end{eqnarray}
If one considers the $qq' \to qq'$ process, then
\begin{eqnarray}
\frac{d\sigma (qq' \to qq')}{dt}
&=&
\frac{4 \pi \alpha_s^2}{9}
\frac{\hat s^2 + \hat u^2}{\hat s^2 \hat t^2}.
\end{eqnarray}
At $\theta_c \sim 90^{\rm o}$, we have $x_a=2x_c$, and we have for $qq'\to q
q'$
\begin{eqnarray}
\frac{d\sigma(qq' \to qq')}{dt}
&=&
\frac{4 \pi \alpha_s^2}{9c_T^4}
\frac{5}{16}.
\label{38}
\end{eqnarray}
In either case, the differential cross section varies as
$d\sigma(ab\to cX')/dt \sim 1/(c_T^2)^2$.

\section{Effects of Multiple Scattering of Partons
on Differential Cross Sections}

Hadrons are composite objects containing a number of partons.  The
collision of hadrons involves the soft and hard collisions of partons.
We separate the total parton-parton cross section $\sigma_{\rm in}$
into soft and hard parts, $\sigma_{\rm in}({\rm
  parton-parton})=\sigma_s + \sigma_H$, where $\sigma_s$ involves soft
processes at low-$p_T$ in the fragmentation of partons in a flux tube
or a string.  The hard cross section $\sigma_H$ involves infrared
singularities at small momentum transfer which can be regulated by a
minimum momentum transfer cutoff $p_0$ that delimits the boundary
between soft and hard processes.  The parton-parton hard cross section
includes the cross section for the production of high-$p_T$ particles
and mini-jets.

With increasing collision energies, we probe regions of smaller $x$, where the parton density increases rapidly.   The number of partons and the total hard-scattering cross section in $pp$ collisions increases with increasing collision energies.  
The total $pp$ hard-scattering cross section may exceed the inelastic
$pp$ total cross section at high energies \cite{Sjo87}.  The average
number of parton-parton interactions above a minimum $p_{0}$ may be
greater than unity.

The presence of a large number of partons in the colliding system leads to parton
multiple scattering in which a projectile parton may make multiple
hard scattering with target partons (also called the rescattering of
partons).  Furthermore, in a hadron-nucleus collision, there are
partons in nucleons along the incident parton trajectory, and multiple
hard scattering of the incident parton with many target partons may
occur.
\begin{figure} [h]
\includegraphics[scale=0.35]{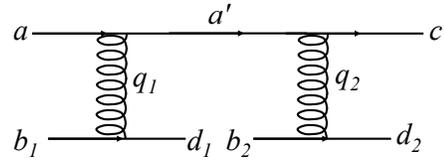}
\caption{The Feynman diagram for the multiple hard scattering process,
  $a+(b_1+b_2) \to c+(d_1+d_2)$, with the exchange of gluons $q_1$ and
  $q_2$.}
\end{figure}

We consider the scattering from an incident parton $a$ to the final
parton $c$ after colliding with two hard scatterers $b_1$, and $b_2$
in the process
\begin{eqnarray}
a + (b_1 + b_2) \to c + (d_1 + d_2)~,
\end{eqnarray}
\noindent
as represented by the Feynman diagram in Fig.\ 1.  For simplicity, we
neglect intrinsic $p_T$ and rest masses so that $\bb a_T=\bb
b_{T1}=\bb b_{T2}=0$.  We are interested in hard-scattering processes
and consider the collision to take place in the center-of-mass system
of $a$ and the partons $(b_1+b_2)$ so that the incident $a$ comes
along the longitudinal $z$ axis and comes out as the final particle $
c $ in the transverse direction at $\theta_c \sim 90^{\rm o}$.  We shall
examine here the influence of the multiple hard-scattering process on
the differential cross section from parton $a$ to parton $c$.

The scattering between $a$ and $b_i$ in Fig.\ 1, with $i=1,2$, is
individually a hard scattering process with the transfer of a
substantial amount of the transverse momentum $\bb q_{Ti}$(=$\bb
d_{Ti}$).  The transverse coherence time $\hbar/(|\bb q_{Ti}|c)$,
which is also the hard-scattering transverse collision time, is quite
short (of the order of 0.01--0.1 fm/c).  On the other hand, at high
energies the total hard-scattering cross section is of order of the
$pp$ inelastic cross section.  The mean-free path $\lambda$ between
parton hard-scattering collisions is of the order of the transverse
radius of the proton.  Therefore, in a multiple hard-scattering
process, the mean-free time $\lambda/ c$ between hard-scattering
collisions is much greater than the transverse hard-collision time
$\hbar/(|\bb q_{Ti}|c)$.

As a consequence, the sequence of hard-scattering collisions of the
incident parton $a$ with scatterers $b_1$ and $b_2$ are incoherent
collisions.  The hard-scattering process $a+b_1 \to a'+d_1$ has been
completed before the other hard-scattering process $a'+b_2 \to c +d_2$
begins.  This implies that the hard-scattering process $a+b_1 \to
a'+d_1$ and the other hard-scattering process $a'+b_2 \to c +d_2$ in
Fig.\ 1 are separately successive two-body hard-scattering processes
with the intermediate particle $a'$ essentially on the mass shell.
These successive hard scatterings can be represented by scattering
laws $d\sigma(ab_1\to a'd_1)/d\bb q_{Ti} \propto \alpha_s^2/(\bb q_{T1}^2)^2$
and $d\sigma(a'b_2\to cd_2)/d\bb q_{T2} \propto \alpha_s^2/(\bb q_{T2}^2)^2$,
with the differential elements $d\bb d_{Ti}=d\bb q_{Ti}$.  The
differential cross section after the multiple
hard-scattering collisions with partons in the other hadron is
therefore
\begin{eqnarray}
&& d \sigma_H^{(2)}(a+(b_1+b_2) \to c+(d_1+d_2))
\nonumber\\
&&\hspace*{1.0cm} \propto  \frac{ d\bb c_T \alpha_s^2 d\bb q_{T1} \alpha_s^2 d\bb q_{T2}}{(\bb q_{T1}^2)^2 
(\bb q_{T2}^2)^2}
\delta (\bb c_T+\bb q_{T1}+\bb q_{T2}),
\label{29a}
\end{eqnarray} 
where the factor $ \alpha_s^4/ [(\bb q_{T1}^2)^2 (\bb q_{T2}^2)^2 ]$ comes from
the the two gluon propagators in Fig.\ 1.  The hard-scattering cross
section from $a$ to $c$, $d\sigma_H^{(2)}(a\to c)$, can be obtained
from the above by integrating over $\bb q_{T1}$ and $\bb q_{T2}$,
regulated by a minimum momentum transfer cutoff $p_0$.

\begin{figure} [h]
\includegraphics[scale=0.40]{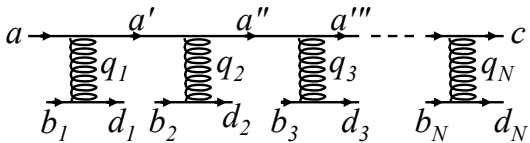}
\caption{The Feynman diagram for the hard scatterings process $a+(b_1+b_2
  +...+b_N) \to c+(d_1+d_2+...+d_N)$, with the exchange of $N$ gluons
  $q_1$, $q_2$, ..., $q_N$.}
\end{figure}

We can generalize the above result for the scattering of the parton
$a$ into the parton $c$ after making a multiple hard scattering with
$N$ hard scatterers as shown in the Feynman diagram in Fig. 2,
\begin{eqnarray}
a + (b_1 + b_2 + ... + b_N)  \to
c + (d_1 + d_2 + ... + d_N).~~~~~~
\label{eq47a}
\end{eqnarray}
Using arguments similar to those leading to Eq.\ (\ref{29a}), the
differential cross section for the multiple hard scattering of $a$ to
$c$ after colliding with $N$ hard scatterers in the other hadron is
\begin{eqnarray}
&&d \sigma_H^{(N)}(a+(b_1+...+b_N)) \to c+(d_1+...+d_N))
\nonumber\\
&&\hspace*{0.9cm} \propto  d\bb c_T  \prod_{i=1}^N\left ( \frac{\alpha_s^2 d \bb q_{Ti} }{(\bb q_{Ti}^2)^2} \right )
\delta (\bb c_T+\bb q_{T1}+...+\bb q_{TN}),~~~~~
\label{eq29}
\end{eqnarray} 
where the factor $\alpha_s^{2N}/ [(\bb q_{T1}^2)^2 ...(\bb q_{TN}^2)^2 ]$ comes
from the $N$ gluon propagators in Fig.\ 2.  The hard-scattering cross
section from $a$ to $c$, $d\sigma_H^{(N)}(a\to c)$, can be obtained
from the above by integrating over $\bb q_{T1}$,..,$\bb q_{TN}$,
regulated by a minimum momentum transfer cutoff of $p_0$.

\section{Effects of  the Multiple Scattering and $T(b)$ on 
the Transverse Differential Cross Section}

The discussions in the last section pertain to the differential cross
section in the scattering of a parton with $N$ parton scatterers.  A
hadron-hadron collision consists of a weighted sum of parton-parton
collision with different number of scatterers $N$, depending on the
transverse profile of the composite target system and the selection of
the centrality of the collision events.

From the earlier studies of multiple hard-scattering processes
\cite{Kas87,Cal90,Cal90a,Cal91,Cal94,Cal94a,Cal01,Acc01,Gyu01,Cor10},
a simple picture emerges to indicate that for high $p_T$ in
minimum-biased events without centrality selection, the sum of the
multiple collision series over different number of scatterers is
dominated by the single scattering differential cross section with the
$1/p_T^4$ dependency.  There are in addition interesting shadowing
cancellations to give logarithmic residual terms.  We would like to
extend the multiple hard-scattering results of \cite{Acc01} to obtain
the explicit power law and logarithmic 
dependence of the multiple scattering 
cross section on target
scatterer number $N$, the dependence on target parton number $A$, as well as on the
centrality of the collision.

The parton-parton hard-scattering cross section $\sigma_{{}_H}$ will shadow 
hard scattering
of the colliding partons.  Thus, for the collision of a parton $a$ on
the object $b$ with $A$ partons, the probability for $N$
hard-scattering collisions at an impact parameter $b$ is \cite{Bla81}
\begin{eqnarray}
P(N,\bb b) = \frac{A!}{N!(A-N)!} [T(b) \sigma_{{}_H}]^N [1-T(b) \sigma_{{}_H}]^{A-N}.~~~~~
\end{eqnarray}
The total hard-scattering cross section for the scattering of $a$ on $N$ partons is 
\begin{eqnarray}
&&\sigma_{H}^{({\rm tot})}(a+A\to cX)
\nonumber\\
&&= \int d\bb b
 \sum_{N=1}^A 
\frac{A!}{N!(A-N)!} [T(b) \sigma_{{}_H}]^N [1-T(b) \sigma_{{}_H}]^{A-N}.~~~
\end{eqnarray}
Thus, the total differential cross section is
\begin{eqnarray}
&&\frac{d \sigma_{H}^{({\rm tot})} (a+A\to cX)}{d\bb c_T}\label{169} \\
&&= \int d\bb b
 \sum_{N=1}^A 
\frac{A!}{N!(A-N)!} [T(b)]^N  \frac{d\sigma_{{}_H}^{(N)}}{d\bb c_T} [1-T(b) \sigma_{{}_H}]^{A-N},
\nonumber
\end{eqnarray}
where the superscript $(N)$ stands for the incident parton making $N$
collisions with target partons.  From Eq.\ (\ref{eq29}), we have
\begin{eqnarray}
\frac{d\sigma_{{}_H}^{(N)}}{d \bb c_T}(\bb c_T)
=\int \prod_{i=1}^{n}\left ( \frac{\alpha_s^2 d\bb q_{iT}}{\bb q_{iT}^4}\right )  \delta(\bb c_T - \sum_{i=1}^N \bb q_{iT}).
\label{eq33}
\end{eqnarray}
In the sum in Eq.\ (\ref{169}), $d\sigma_{{}_H}^{(N)}/d \bb c_T$ is
of order $\alpha_s^{2N}$.  The absorption part is represented by the
term $(1-T\sigma_H)^{A-N}$. We can expand the absorption part $[1-T(b)
  \sigma_{{}_H}]^{A-N}$ as a power series, and we obtain
\begin{eqnarray}
&&\frac{d \sigma_{H}^{({\rm tot})} (a+A\to cX)}{d\bb c_T}\
\nonumber\\
&= &\int d\bb b
A T(b)  \frac{d\sigma_{{}_H}^{(1)}}{d\bb c_T}\biggl \{ \{1-(A-1)[T(b)\sigma_H]
\nonumber\\
&& \hspace*{3.4cm}+\frac{(A-1)(A-2)}{2}[T(b)\sigma_H]^2\biggr  \}
\nonumber\\
&+& \int d\bb b
\frac{A(A+1)/}{2} [T(b)]^2  \frac{d\sigma_{{}_H}^{(2)}}{d\bb c_T}\biggl  \{1-(A-2)[T(b)\sigma_H]\biggr  \}
\nonumber\\
&+&\int d\bb b
 \frac{A(A+1)(A+2)}{6}  [T(b)]^3  \frac{d\sigma_{{}_H}^{(3)}}{d\bb c_T}+ ...
\end{eqnarray}
After expanding the absorption term $[1-T(b) \sigma_{{}_H}]^{A-N}$, we
can collect all terms of the same order in $\alpha_s^{2N}$ to resum
Eq.\ (\ref{169}) in the form
\begin{eqnarray}
&&\frac{d \sigma_{H}^{({\rm tot})} (a+A\to cX)}{d\bb c_T}
\nonumber\\
&& \hspace*{1.6cm}= \int d\bb b
 \sum_{n=1}^N 
\frac{A!}{A!(A-N)!} [T(b)]^N  \frac{d\tilde \sigma_{{}_H}^{(N)}}{d\bb c_T},
\end{eqnarray}
where $ {d\tilde \sigma_{{}_H}^{(N)}}/{d\bb c_T}$ is of order
$\alpha_s^{2N}$ given by
\begin{subequations}
\label{eq36}
\begin{eqnarray}
 \frac{d\tilde \sigma_{{}_H}^{(1)}}{d\bb c_T}
&=&\frac{d \sigma_{{}_H}^{(1)}}{d\bb c_T}\label{eq36a} \\
 \frac{d\tilde \sigma_{{}_H}^{(2)}}{d\bb c_T}
&=&\frac{d \sigma_{{}_H}^{(2)}}{d\bb c_T}-\frac{2(A-1)}{A+1}
\frac{d \sigma_{{}_H}^{(1)}}{d\bb c_T}\sigma_H
\label{eq36b}\\
 \frac{d\tilde \sigma_{{}_H}^{(3)}}{d\bb c_T}
&=&\frac{d \sigma_{{}_H}^{(3)}}{d\bb c_T}
-\frac{3(A-2)}{(A+2)}
\frac{d \sigma_{{}_H}^{(2)}}{d\bb c_T}\sigma_H
\nonumber\\
& &~~~~~~~~+\frac{3(A-1)(A-2)}{(A+1)(A+2)}
\frac{d \sigma_{{}_H}^{(1)}}{d\bb c_T}\sigma_H^2.
\label{eq36c}
\end{eqnarray}
\end{subequations}
The last term in Eq.\ (\ref{eq36b}) and the last terms in
Eq.\ (\ref{eq36c}) represent shadowing corrections due to the
absorption factor $[1-T(b) \sigma_{{}_H}]^{A-N}$.  The basic
parton-parton collision gives
\begin{eqnarray}
\frac{d\tilde \sigma_{{}_H}^{(1)}}{d  \bb c_T}(\bb c_T)
&\sim& \frac{\alpha_s^2}{\bb c_T^4},
\label{eq39}
\end{eqnarray}
where for simplicity a constant coefficient that depends on the nature
of the partons as in Eqs.\ (\ref{36}) and (\ref{38}) has been
understood.  The integrated cross section with a cutoff at $p_{0}$
gives
\begin{eqnarray}
\sigma_{{}_H}^{(1)}& \sim& 
 \frac{\pi  \alpha_s^2}{p_0^2}.
\end{eqnarray}
We consider the case with $A\gg 1$ in Eq.\ (\ref{eq36}), and we obtain
\begin{eqnarray}
&&\frac{d \tilde \sigma_{H}^{(2)} (a\to c)}{d \bb c_T}
\nonumber\\
&&=
2 \left \{ \alpha_s^4\int_{p_0}^{c_T/2} \left ( \frac{ d\bb q_{1T} }
{\bb q_{1T}^4 (\bb c_T-\bb q_{1T})^4}\right ) 
-\frac{d\sigma_{{}_H}^{(1)}}{d \bb c_T} \sigma_{{}_H}\right \}.
\end{eqnarray}
In the integration in the above sum, the dominant contribution comes
from the region around $q_{1T}\sim 0$.  We expand $1/(\bb c_T-\bb
q_{1T})^4$ about $q_{1T}\sim 0$.  As a result of the shadowing
cancellation in Eq.\ (\ref{eq36b}) or (\ref{eq39}), the singular terms
proportional to $1/p_0^6$ cancel out and only a logarithmic term
remains \cite{Acc01}.  We find that $d \tilde \sigma_{H}^{(2)} /{d \bb c_T}$ is given explicitly by 
\begin{eqnarray}
\frac{d \tilde \sigma_{H}^{(2)} (a\to c)}{d \bb c_T}
&=&
\frac{16\pi\alpha_s^4}{c_T^6}
\ln \{\frac{c_T}{2p_0}\},
\label{eq42}
\end{eqnarray}
which has a power law $1/c_T^6$ multiplied by a mild logarithm term.  
Next, we need to study $N=3$,
\begin{eqnarray}
&&\frac{d \tilde \sigma_{H}^{(3)} (a\to c)}{d \bb c_T}
\nonumber\\
&&=
\left \{ \frac{d\sigma_{{}_H}^{(3)}}{d \bb c_T} 
- 3 \frac{d\sigma_{{}_H}^{(1)}}{d \bb c_T} \sigma_{{}_H}^2\right \}
- 3 \left \{ \frac{d\sigma_{{}_H}^{(2)}}{d \bb c_T}
-2 \frac{d\sigma_{{}_H}^{(1)}}{d \bb c_T}\sigma_{{}_H}
 \right \}
 \sigma_{{}_H}.~~~~~
\label{182}
\end{eqnarray}
We expand $1/(\bb c_T-\bb q_{iT})^4$ again about $q_{iT}\sim 0$.
Similarly, the singular terms proportional to$1/p_0^8$ cancel out, and
only the logarithmic term remains.  We find  that  $d \tilde \sigma_{H}^{(3)}/d \bb c_T$ is given by
\begin{eqnarray}
\frac{d \tilde \sigma_{H}^{(3)} (a\to c)}{d \bb c_T}
=\frac{3\pi^2\alpha_s^6 }{c_T^8}\times 312[ \ln \frac{c_T}{3p_0}]^2.
\label{eq44}
\end{eqnarray}
Equations (\ref{eq39}), (\ref{eq42}), (\ref{eq44}) give explicitly the differential cross sections of a parton after
multiple scattering with $N$ scatterer partons as
\begin{eqnarray}
\frac{d \tilde \sigma_{H}^{(N)} (a\to c)}{d \bb c_T}\propto
\frac{\alpha_s^{2N}}{c_T^{2+2N}}[ \ln \frac{c_T}{Np_0}]^{N-1} ,
\end{eqnarray}
which states that  the differential cross section for
multiple parton scattering  obeys a power laws with the power index  
(2+2$N$), multiplied by a logarithm function $[\ln(c_T/Np_0)]^{N-1}$.  For the scattering of a parton with
one scatterer, it gives $\alpha_s^2/p_T^4$, with two scatterers it
gives $\alpha_s^4 \ln (p_T/2p_0)/p_T^6$, and with three scatterers it
gives $\alpha_s^6[\ln (p_T/3p_0)]^2/p_T^8$.

Collecting the terms together, we obtain the differential cross
section for the collision of a parton with a composite  system
with $A$ partons and a thickness function $T(b)$  given by
\begin{eqnarray}
& &\frac{d \sigma_{H}^{(tot)} (a\to c)}{d \bb c_T}
=  A \frac{\alpha_s^2}{ c_T^4} \int d\bb b ~T(b)
\\
& &\hspace*{1.0cm}
+
\frac{A(A-1)}{2}
\frac{16\pi\alpha_s^4}{c_T^6}
\ln \{\frac{c_T}{2p_0}\}
\int d\bb b[T(b)]^2 
\nonumber\\
& &\hspace*{1.0cm}
+
\frac{A(A-1)(A-2)}{6}
\frac{936\pi^2\alpha_s^6 }{c_T^8}[ \ln \frac{c_T}{3p_0}]^2
\int d\bb b[T(b)]^3. \nonumber
\label{eq46}
\end{eqnarray}
Depending on the limits of the impact parameter integration, the above
result gives the differential cross section for collisions with
different centrality selections.  For minimum-biased events without an
impact parameter selection, one sums over the whole range of impact
parameters.  We can consider a thickness function $T(b)$ in the form
of a Gaussian given by \cite{Won94}
\begin{eqnarray}
T(b)=\frac{\exp\{-b^2/2\beta^2\}}{2\pi \beta^2},
\end{eqnarray}
where $\beta=r_0 /\sqrt{3}$.  [For a proton, $r_0\sim$ 0.7 fm
  \cite{Won94}].  We then have
\begin{eqnarray}
\int d \bb b [T(b) ]^N
&=& 
\frac{1}{N (2\pi\beta^2)^{N-1}},
\end{eqnarray}
and the minimum-biased differential cross section is 
\begin{eqnarray}
\frac{d \sigma_{H}^{(tot)} (a\to c)}{d \bb c_T}&
=&  
A  \frac{\alpha_s^2}{ c_T^4}
\nonumber\\
& & \hspace*{-2.0cm}
+\frac{A(A-1)}{2}
\frac{1}{2(2\pi\beta^2)}
\frac{16\pi\alpha_s^4}{c_T^6}
\ln \{\frac{c_T/2}{p_0}\}
\nonumber\\
& & \hspace*{-2.0cm}
+\frac{A(A-1)(A-2)}{6}
\frac{312}{3(2\pi\beta^2)^2}
\frac{3\pi^2\alpha_s^6 }{c_T^8}[ \ln \frac{c_T}{3p_0}]^2.~~~~
\label{eq47}
\end{eqnarray}
For another sharp-cutoff thickness function $T(b)$ given by \cite{Won94}
\begin{eqnarray}
T(b)=\frac{3}{2\pi R^3}\sqrt{R^2-b^2}~ \Theta(R-b),
\end{eqnarray}
we obtain
\begin{eqnarray}
\int d\bb b [T(b)]^N 
=\frac{3^N}{(N+2)2^{N-1}\pi^{N-1}R^{2N-2}},
\end{eqnarray}
and the  minimum-biased differential cross section is 
\begin{eqnarray}
\frac{d \sigma_{H}^{(tot)} (a\to c)}{d \bb c_T}
&=&
A  \frac{\alpha_s^2}{ c_T^4}
\nonumber\\
& &  \hspace*{-2.0cm}
+
\frac{A(A-1)}{2}
\frac{3^2}{4\times 2\pi R^2}
\frac{16\pi\alpha_s^4}{c_T^6}
\ln \{\frac{c_T/2}{p_0}\}
\nonumber\\
& &  \hspace*{-2.0cm}
+
\frac{A(A-1)(A-2)}{6}
\frac{3^3312}{5\times 2^2\pi^2 R^4}
\frac{3\pi^2\alpha_s^6 }{c_T^8}[ \ln \frac{c_T}{3p_0}]^2.~~~~~~
\label{eq50a}
\end{eqnarray}
Because the power index increases with $N$ as $2+2N$, the minimum-biased differential
cross section at high $c_T$ in Eq.\ (\ref{eq47}) or (\ref{eq50a}) will be  dominated
by the differential cross section for a single parton-parton $N$=1 collision,
varying as $\alpha_s^2/p_T^4$.

It should however be recognized that even though the lowest order $\alpha_s/c_T^4$ dominates at the highest $c_T$ region,  contributions  higher order in $\alpha_s$ begin to enter into play under certain circumstances.   For example, as the transverse momentum is lowered below the highest $c_T$ region, there will be values of $c_T$ when contributions with higher power index such as $1/c_T^6$ and $1/c_T^8$ in the above series in Eq.\ (\ref{eq47}) or Eq.\ (\ref{eq50a})
begin to be important, depending on the value of $A, \beta$(or $R$), and  $\alpha_s$.   In another example,  as the cone radius $R$ in jet measurements increases, the cone region will contain parton-parton processes with a greater number of interacting vertices, and it may become necessary to include higher and higher order contributions where  contributions of order $\alpha_s^{2N}$ arising from multiple scattering will have a power index $2+2N$.  

We note in passing that Eq.\ (\ref{eq46}) also gives the centrality
dependence of the differential cross section,
\begin{eqnarray}
& &\frac{d \sigma_{H}^{(tot)} (a\to c)}{d  \bb c_T^2d \bb b}( c_T, \bb b)
=  A \frac{\alpha_s^2}{ c_T^4}~T(b)
\\
& &\hspace*{1.0cm}
+
\frac{A(A-1)}{2}
\frac{16\pi\alpha_s^4}{c_T^6}
\ln \{\frac{c_T}{2p_0}\}
 [T(b)]^2 
\nonumber\\
& &\hspace*{1.0cm}
+
\frac{A(A-1)(A-2)}{6}
\frac{936\pi^2\alpha_s^6 }{c_T^8}[ \ln \frac{c_T}{3p_0}]^2
 [T(b)]^3. \nonumber
\label{eq51}
\end{eqnarray}
The above result indicates that one can alter the weights of the
different number of scatterers and the power index $n$, by an impact
parameter selection.  The number of partons $A$ in a hadron or a
nucleus is a dynamical quantity that may depend on the probing
transverse momentum and the target nucleus mass number, and it is not
yet a well-determined quantity.  It is an interesting experimental
question whether the numbers of partons $A$ may be so large in some
phase space regions or some collision energies as to make it possible to alter the power law
behavior of the transverse differential cross section for selected
centralities, over different $p_T$ regions.   One expects that as the centrality becomes more and more central, contributions with a greater number of multiple parton collisions gains in importance.  As a consequence, the power index $n$ is expected to become greater when we select more central collisions.   

\section{Comparison of Relativistic Hard-Scattering Model 
with Experimental Jet Transverse 
differential cross sections}

The results in the last section show that without centrality
selection in minimum-biased events, the differential cross section for
the production of partons at high-$p_T$ will be  dominated by the contribution from a single parton-parton scattering that behaves as
$1/c_T^4$,
 \begin{eqnarray}
\frac{d \sigma_{H}^{(tot)} (a\to c)}{d \bb c_T}
\propto  \frac{\alpha_s^2}{ c_T^4},
\label{52}
\end{eqnarray}
in line with previous analyses on the multiple scattering process in
\cite{Kas87,Cal90,Cal90a,Cal91,Cal94,Cal94a,Cal01,Acc01,Gyu01,Cor10}.
Multiple scatterings with $N>1$ scatterers contribute to terms of
order $\alpha_s^{2N}$ and involve a power law
$[\ln{(C_T/Np_0)}]^{N-1}/c_T^{2+2N}$.

We now consider the lowest order result of Eq.\ (\ref{52}).
From Eqs.\ (\ref{rhs}) and (\ref{52}), the relativistic hard
scattering cross section of Eq.\ (\ref{rhs}) for the collision of
hadrons $A$ and $B$ when a parton $a$ of one of the hadron makes a
hard scattering with a partons in the other hadron to produce the
parton $c$ is
\begin{eqnarray}
&&E_p \frac{d^3\sigma( AB \to c X) }{dc^3}
= \frac{d^3\sigma( AB \to c X) }{dy d{\bb c}_T}
\nonumber\\
&&\hspace*{1.0cm}
\propto \frac{\alpha_s^2(Q^2(c_T^2))(1\!-\!x_{a0}(c_T))^{g_a}(1\!-\!x_{b0}(c_T))^{g_a} }
{c_T^{4} [c_T/\sqrt{s}]^{1/2}}.~~~~~~
\label{AB}
\end{eqnarray}
Different factors in the above equation (\ref{AB}) reveal the physical
origins and the associated degrees of freedom.  The power law
$\alpha_s^2/c_T^{4}$ arises from parton-parton hard scattering.  The additional
$c_T^{1/2}$ in the denominator comes from the $1/\sqrt{\tau_c}$ factor
in Eq.\ (\ref{rhs}) and it arises from the integration of the momentum
fraction of the other colliding parton $x_b$.  The structure function
factor $(1-x_{a0}(c_T))^{g_a}(1-x_{b0}( c_T))^{g_a}$ comes from the
probability for the occurrence of the momentum fractions of the
colliding partons.  The quantities $x_{a0}(c_T)$ and $x_{b0}(c_T)$ are
functions of $c_T$ as given in Eqs.\ (\ref{44}) and (\ref{44a})
respectively.  The argument $ c_T$ inside the structure function
factor is the transverse momentum of the scattered parton $c$, prior
to its fragmentation.  The exponential indices $g_a$ and $g_b$ come from the structure functions.  They  can also be
estimated from the spectator counting rule of Blankenbecler and
Brodsky \cite{Bla74} as given by $g_{\{a,b\}}=2n_s-1$, where $n_s$ is
the number of spectators of the composite hadron system $a$ or $b$ in
the hard-scattering collision.  This is essentially the form of the
cross section as first suggested by Blankenbecler, Brodsky, and
collaborators \cite{Bla74,Bla75,Sch77,Bla77,Won94}.

\begin{figure} [h]
\includegraphics[scale=0.45]{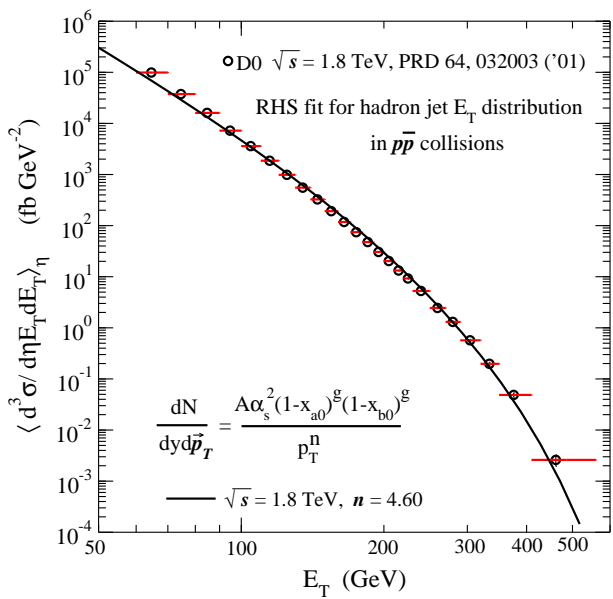}
\caption{(Color online) Comparison of the experimental $d\sigma/d\eta
  E_T dE_T$ data from the D0 collaboration \cite{Abb01} for the
  distribution of hadron jet transverse energy $E_T$ at
  $|\eta|$$<$0.5, in $p\bar p$ collision at $\sqrt{s}$=1.8 TeV, with
  the relativistic hard-scattering model result in Eq.\ (\ref{ABp}).}
\end{figure}

The results of Eq.\ (\ref{AB}) can be compared directly with the
transverse differential cross sections for hadron jet and isolated
photon production.  Previously, Arleo $et~al.$ \cite{Arl10} have
presented a method to obtain an ``experimental" local power index
$n^{\rm exp}(x_c)$.
  Specifically, referring to
Eq.\ (\ref{AB}) and representing the power index of $c_T$ by $n$, 
 the  lowest order theoretical
result of Eq.\ (\ref{AB})  predicts $n$=4+1/2. 
One 
focuses attention at a fixed $x_c$$(=c_T/\sqrt{s})$ at $\eta=0$ for which $x_{a0}=x_{b0}=
2x_c$.   
Upon neglecting the $\sqrt{s}$ dependence of $\alpha_s^2$, one  extracts an experimental  power index $n(x_c)$ as a function of $x_c$ 
 by comparing the invariant cross sections at a
fixed $x_c$ at different collision
energies, 
  \cite{Arl10}
\begin{eqnarray}
n(x_c)\sim \frac{\ln [\sigma_{\rm inv}(\sqrt{s_1},x_c)/
{\sigma_{\rm inv}(\sqrt{s_2},x_c)}]}
{\ln \left [{\sqrt{s_2}}/{\sqrt{s_1}}\right ] } +\frac{1}{2},
\end{eqnarray}
which is  related to the quantity $n^{\rm exp}(x_c)$ of Arleo $et~al.$ \cite{Arl10} by
$n(x_c)$=$n^{\rm exp}(x_c)$+1/2.    Table I summarizes the
average experimental power index $\langle n^{\rm exp}\rangle$
extracted by Arleo $et~al.$ \cite{Arl10} from the D0 and CDF photon and hadron jet
transverse differential cross sections
\cite{Abe93,Abb01,Aco02,Abb00,Aba01}.  The power indices have the
values of $\langle n\rangle$=$\langle n^{\rm exp}\rangle$+1/2=4.8--5.2.  The local power
indices as a function of $x_c$ are also shown in Fig. 2 of Arleo $et~al.$
\cite{Arl10}.  These power indices are in approximate agreement with the power index 
$n$=4.5 in Eq.\ (\ref{AB}) obtained in the relativistic hard-scattering
model in perturbative QCD.

\begin{figure} [h]
\includegraphics[scale=0.45]{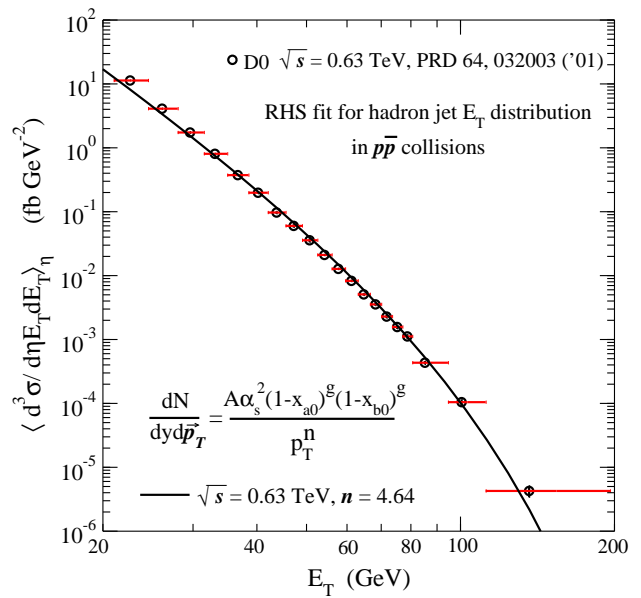}
\caption{(Color online) Comparison of the experimental $d\sigma/d\eta
  E_T dE_T$ data from the D0 collaboration \cite{Abb01} for the
  distribution of hadron jet transverse energy $E_T$ at $|\eta|$$<$0.5
  , in $p\bar p$ collision at $\sqrt{s}$=0.630 TeV, with the
  relativistic hard-scattering model result in Eq.\ (\ref{ABp}).}
\end{figure}

\vspace*{0.3cm} 

\begin{table}[h]
\caption { The mean power index $\langle n^{\rm exp}\rangle$ extracted
  from experimental transverse differential cross sections for hadron
  and photon jet productions in $p\bar p$ collisions
  at Fermilab as obtained in \cite{Arl10} by comparing the invariant
  cross sections at different energies.  }
\vspace*{0.3cm}
\begin{tabular}{|c|c|c|c|c|}
\cline{1-5}
      Collaboration  & Ref.            & Particles &  $\sqrt{s}$  & $\langle n^{\rm exp}\rangle $   \\ 
                    &    &   $  $ &  (TeV)  &    \\ 
 \hline
CDF& \cite{Abe93} & hadrons   &   0.546,~1.8 &  4.3$\pm$0.09  \\ \hline
D0   &\cite{ Abb01} & hadrons  &   0.630,~1.8&  4.5$\pm$0.04   \\ \hline
CDF  & \cite{Aco02,Abb00}& photons  &  0.630,~1.8  &  4.7$\pm$ 0.09   \\ \hline
D0  & \cite{Aba01} & photons  &  0.630,~1.8  &  4.5$\pm$ 0.12   \\ \hline
\end{tabular}
\end{table}
As an example to provide a complementary comparison, we focus our
attention at a fixed collision energy and express the differential jet
cross section $d^3\sigma( AB \to p X) /dy d{\bb c}_T$ in
Eq.\ (\ref{AB}) as
\begin{eqnarray}
&& \frac{d^3\sigma( AB \to c X) }{dy d{\bb c}_T}
\nonumber\\
&&\hspace*{0.7cm} =A \frac{\alpha_s^2(Q^2(c_T))(1\!-\!x_{a0}(c_T))^{g_a}(1\!-\!x_{b0}( c_T))^{g_a} }
{c_T^{n}}.~~~~~~
\label{ABp}
\end{eqnarray}
where $c_T$$\sim$$E_T$, $d\bb c_T=2\pi E_T dE_T$.
 We also use the symbol $p_T$ for the jet transverse momentum $c_T$.
The coupling constant $\alpha_s$ is a function of $Q^2$, which will be identified as $p_T^2$.  We use the running QCD coupling constant
\cite{Ber12}
\begin{eqnarray}
\alpha_s(p_T) = \frac{12\pi}{27 \ln(p_T^2/\Lambda_{\rm QCD}^2)},
\end{eqnarray}
where  $\Lambda_{\rm QCD}$=0.25 GeV has been chosen  such that
$\alpha_s(M_Z^2)=0.1184$.
We infer from Eq.\ (\ref{ABp}) 
\begin{eqnarray}
&&n = - \frac{d}{d\log p_T} \bigg \{  \log \frac{d\sigma}{d\eta \, p_T dp_T}
\nonumber\\
&&\hspace*{0.7cm}
- \log\bigl [ \alpha_s^2(p_T)(1\!-\!x_{a0}(p_T))^{g_a}(1\!-\!x_{b0}( c_T))^{g_a} \bigr]
 \biggr \}.~~~~~~~
\end{eqnarray}
In the region where $p_T \ll
\sqrt{s}$ and the variation of $\alpha_s$ with $p_T$ is not large, the quantity $\log (d\sigma/dy\, p_Tdp_T)$ will be  approximately a linear
function of $\log p_T$.   The log-log plot of $\log (d\sigma/dy\,
p_Tdp_T)$ as a function of $\log p_T$ should appear nearly as a straight
line, with the power index $n$ given by the magnitude of slope of the line.    In Figs. 3 and 4,  
the straight lines  in the lower $E_T$ regions  
exhibit such a linear behavior.

We use Eq.\ (\ref{ABp}) to search for the parameters $A$ and $n$ to
fit the hadron jet transverse 
differential cross section as a function of 
$E_T$($\sim p_T$)  at $\eta\sim 0$ in $p\bar
p$ collisions at Fermilab.  The exponential index $g_a=g_b$ for the
structure function of a gluon varies from 6 to 10 in different
structure functions \cite{Duk84,Che03,Che05}.  We shall take $g_a=6$
from \cite{Duk84}.  The experimental D0 hadron jet data of
$d\sigma/d\eta E_T dE_T$ at $|\eta|$$<$0.5 for $p\bar p$ collision at
$\sqrt{s}$=1.8 TeV \cite{Abb01} can be fitted with 
$n$=4.60 and $2\pi
A$=2.29$\times 10^{15}$ fbGeV$^{-2}$, as shown in
Fig.\ {3}.  The experimental D0 hadron jet data of $d\sigma/d\eta E_T
dE_T$ for $p\bar p$ collision at $\sqrt{s}$=0.630 TeV \cite{Abb01} can
be fitted with 
$n$=4.64 and $2 \pi A$=1.64$\times 10^{9}$ fbGeV$^{-2}$, as shown in Fig.\ {4}.  These power indices are in
approximate agreement with the value of $n$=4.5 in
Eq.\ (\ref{AB}), indicating the approximate validity of the
hard-scattering model description for jet production in hadron-hadron
collisions, with the predominant $\alpha_s^2/c_T^4$ parton-parton differential cross section.  These power indices extracted from the differential cross section are also in approximate agreement with those in Table I extracted by comparing cross sections at two different energies \cite{Arl10}.

In another comparison of the jet production data with the
hard-scattering model, we examine in Fig.\ 5 the jet differential
cross section $d\sigma/d\eta \, p_T dp_T$ in $pp$ collisions at
$\sqrt{s}=2.76$ TeV at the LHC obtained by the ALICE collaboration at
$\eta<0.5$ with $R$=0.4 and 0.2 \cite{Alice13}.  The log-log plot of $\log
[d\sigma/d\eta \, p_T dp_T]$ versus $\log p_T$ gives nearly a straight
line with the slope $-n$. The jet differential cross section can be
fitted with the power index 
$n$=5.0$\pm$0.2 and an overall magnitude
of $2\pi A$=2080 mb GeV$^{-2}$ for $R=0.4$ and  
$n$=4.8$\pm$0.2 and an overall magnitude
of $2\pi A$=535 mb GeV$^{-2}$.  
These power indices are close to the value
of $n=4.5$ expected in Eq.\ (\ref{AB}) in the hard-scattering model.

In another comparison, we show in Fig.\ 6 the jet differential
cross section $d\sigma/d\eta \, p_T dp_T$ in $pp$ collisions at
$\sqrt{s}=7$ TeV at LHC obtained by the CMS collaboration at
$\eta<0.5$ with $R=0.5$ \cite{CMS11}.  
The jet differential cross section can be
fitted with the power index $n$=5.44$\pm 0.1$ and 
of $2\pi A=5.05\times$10$^{14}$ mb GeV$^{-2}$ as shown in Fig.\ 7. 
The value of $n$ is slightly greater than the expected value of $n=4.5$.

\begin{figure} [h]
\includegraphics[scale=0.45]{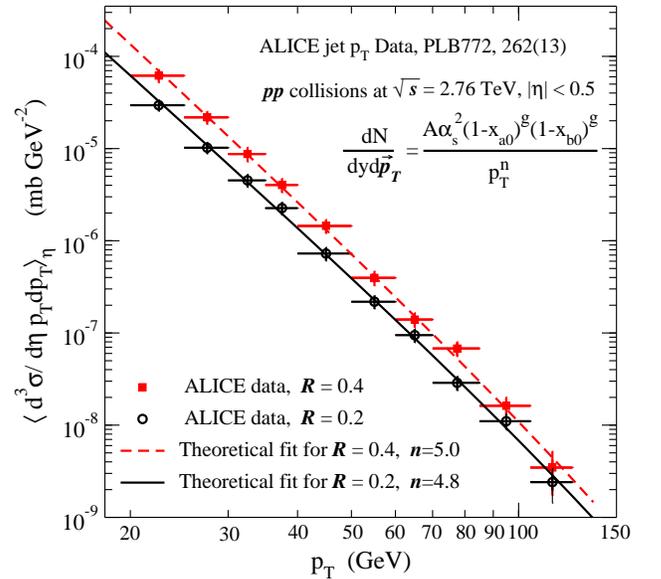}
\caption{(Color online) Comparison of the experimental $d\sigma/d\eta
  p_T dp_T$ data from the ALICE collaboration \cite{Alice13} for the
  transverse momentum distribution of hadron jets
  $d\sigma/d\eta\,p_Tdp_T$ at $|\eta|$$<$0.5, in $p p$ collision at
  $\sqrt{s}$=2.76 TeV, with the relativistic hard-scattering model
  result in Eq.\ (\ref{ABp}). }
\end{figure}

We note  in the last few examples 
that
the power index $n$ increases slightly as the cone radius $R$ increases.  An increase in the cone radius allows the sampling of events with a greater number of  the parton-parton interaction vertices 
inside the cone, and each interaction vertex brings in a power of $\alpha_s^2$.   A greater cone radius has greater contributions from processes that are higher order in $\alpha_s$.  Thus, among  many high next-to-leading order NLO and next-to-next-leading order NNLO contributions, some of the $\alpha_s^4/p_T^6$ contributions of the multiple scattering processes 
discussed in Eqs.\ (\ref{eq47}) and (\ref{eq50a})
in Secs. III and IV may also need to be included.  
Because of the limited number of cases, more measurements will be needed to confirm whether the increase in the power index with increasing  $R$  is a general phenomenon.

We conclude from these comparisons of the transverse differential
cross sections of hadron jets in both $p\bar p$ and $pp$ collisions at
high energies that the data supports the relativistic hard-scattering
description of the collision process, with a basic parton-parton
differential cross section behaving approximately as $\alpha_s^2/p_T^4$
with some tentative evidence of an increase in the power index as $R$ increases.

\begin{figure} [h]
\includegraphics[scale=0.45]{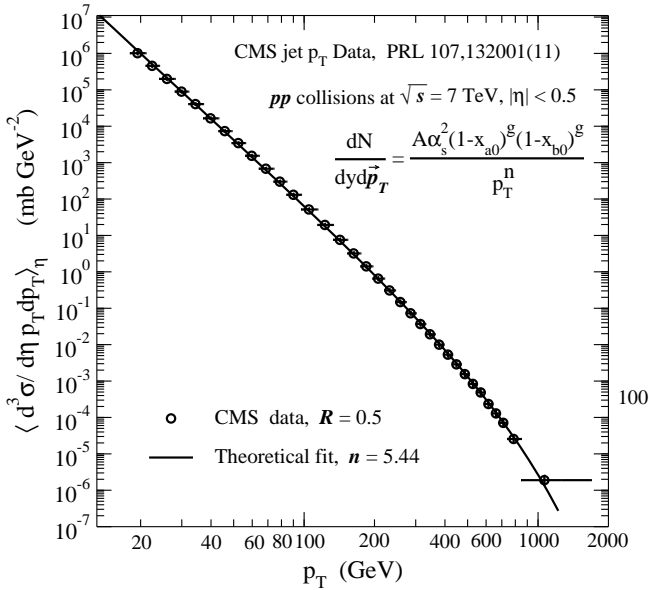}
\caption{(Color online) Comparison of the experimental $d\sigma/d\eta
  p_T dp_T$ data from the CMS Collaboration \cite{cms11} for the
  transverse momentum distribution of hadron jets
  $d\sigma/d\eta\,p_Tdp_T$ at $|\eta|$$<$0.5, in $p p$ collision at
  $\sqrt{s}$=7 TeV, with the relativistic hard-scattering model
  result in Eq.\ (\ref{ABp}). }
\end{figure}

It is interesting to note that when the   structure
function information is known  from other measurements, the
hadron jet spectra differential cross section can be reasonably
described with only a single power index $n$ and an overall magnitude
parameter $A$.  The shape of the transverse differential cross section
of hadron jets has only a very small number of the  degrees of
freedom.

\section{Effects of Fragmentation on 
Transverse Differential Cross Section}

Experimentally, we detect hadrons and the construction of a hadron jet
is inferred from a correlated cone of hadrons.  Experimental
measurements also give hadron spectra at high transverse momenta
without reconstructing jets.  The analyses of the power indices $n$
give $n$$\sim$7 for hadron transverse spectra \cite{Won12} but
$n$=4.5--5 for jet transverse differential cross sections as shown in
the last section.  The difference between the power indices is likely
to arise from the subsequent evolution of the parton.

Other pieces of evidence that the parton-to-hadron final-state
evolution may lead to a change in the power index show up when we
compare the experimental local power indices for jets and for
hadrons \cite{Arl10}. In Fig.\ 2 of Ref.\ \cite{Arl10}, the local
power indices $n^{\rm exp}$ as a function of $x_\perp$=$2x_c$ cluster
around $n^{\rm exp}$$\sim$4.5 for jets but $n^{\rm exp}$$\sim$5--9 for
hadrons.  Furthermore, Table I of \cite{Arl10} gives $\langle n^{\rm
  exp}\rangle$ for jets that are substantially smaller than $\langle
n^{\rm exp}\rangle $ for hadrons.  We need to consider the difference
between jets and hadrons and the fragmentation and showering of jets
(representing partons) to become hadrons.

We shall view the parton fragmentation and the accompanying showering
as equivalently final-state processes and speak of them
interchangeably to emphasize different aspects of the parton
final-state evolution.  In the remaining sections, we shall
consistently use the symbol $c$ to label a parton and its momentum and
the symbol $p$ to label a hadron and its momentum.

In the showering of a parton $c$, a large number of hadrons comes out
nearly collinearly with the parton $c$ in a cone along the ${\bb c}$
direction.  In the present study of high-$p_T$ particles in the
central rapidity region, the parton $c$ is predominantly along the
transverse direction, and the shower of the produced hadrons will also
be along the transverse direction. For the study of the high-$p_T$
spectra as a result of the showering of a parton $c$, it suffices to
focus attention on the leading hadron $p$ of the cone of shower
particles, because of the rapid falloff of the transverse momentum
distribution as a function of increasing $c_T$.  The leading hadron
fragment with transverse momentum $p_T$ contributes significantly to
the final spectra at that $p_T$ whereas nonleading hadron fragments
of the shower contribute only insignificantly to the spectra at their
corresponding $p_T$ values.  Thus, for the examination of the
high-$p_T$ hadron spectra after parton fragmentation and showering,
each parent parton $c$ with a momentum $c_T$ can be viewed as
fragmenting into a single leading hadron $p$ with momentum $p_T$ by
the showering process.

The showering of the partons will go over many generations of
branching and each branching will degrade the momentum of the
showering parton by a momentum fraction $\zeta$.  We can consider the
transverse momentum $p_T$ of the leading hadron as arising from the
$\lambda$th branching generation of the shower.  The 4-momentum of
the leading hadron $p$ and the 4-momentum of the parent parton $c$ are
related by
\begin{eqnarray}
p = \zeta^\lambda c .
\label{eq52}
\end{eqnarray}
We relabel   the cumulative product $\zeta^\lambda$ by the momentum fraction $z$,
\begin{eqnarray}
z=\zeta^\lambda
\end{eqnarray}
to relate $p$ with $c$
\begin{eqnarray}
p=z c.
\label{eq56}
\end{eqnarray}
The probability for the fragmentation of the parton $c$ into the
hadron $p$ is specified phenomenologically by the fragmentation
function $D_{p/c}(z)$, which depends on the QCD momentum transfer
scale.

We shall consider first the simplest case of showering and
fragmentation in which the momentum fraction $z$ is independent of the
magnitude of the parton transverse momentum $c_T$.  We shall consider
more sophisticated showering algorithm in the next section.  In this
case with $z$ independent of $c_T$, the hadron transverse momentum
$p_T$ is a linear function of the parton transverse momentum $c_T$ in
Eq.\ (\ref{eq56}).  Under the fragmentation from the parton $c$ to the
hadron $p$, the differential cross sections ${d\sigma (AB\to pX)
}/{dp^4}$ and ${d\sigma (AB\to cX) }/{dc^4}$ are related by
\begin{eqnarray}
&&\!\!\!\!\frac{d\sigma (AB\to pX) }{dp^4}\nonumber\\
&&=\int dz D_{p/c}(z) \int dc^4 \frac{d\sigma (AB\to cX) }{dc^4}\delta^{(4)}(p - z c).
\end{eqnarray}
We therefore have
\begin{eqnarray}
E_p \frac{d\sigma (AB\to pX) }{dp^3}
&=& \frac{d\sigma (AB\to pX) }{dy d\bb p_T}
\label{eqA61}\\
&&\hspace*{-2.5cm}\propto  \int \frac{dz}{z^2} D_{p/c} (z) z^{4+1/2}
\nonumber\\
&&\hspace*{-1.9cm}
 \times \frac{\alpha_s^2(c_T)(1-x_{a0}(c_T))^{g_a}(1-x_{b0}(c_T))^{g_a} }
{p_T^{4+1/2}},\nonumber
\end{eqnarray}
where the arguments $c_T$ in $x_{a0}$ and $x_{b0}$ are evaluated at
$c_T=p_T/z$.  We can expand the
factor \break $\alpha_s(c_T)(1-x_{a0}(c_T))^{g_a}$$(1-x_{b0}(c_T))^{g_a}$ about $\bar c_T$
in the above equation as a power series of $c_T$,
\begin{eqnarray}
f(c_T)&=& \alpha_s(c_T)((1-x_{a0}(c_T))^{g_a} (1-x_{b0}(c_T))^{g_b} 
\nonumber\\
&=&f(\bar c_T)+(c_T-\bar c_T) f'(c_T)+\frac{(c_T-\bar c_T)^2}{2}  f''(c_T). ~~~~~
\end{eqnarray}
The error in the first order is minimized if $\bar c_T$ is defined as
\begin{eqnarray}
\bar c_T=\langle \frac{p_T}{z} \rangle =p_T\langle \frac{1}{z} \rangle
\end{eqnarray}
where
\begin{eqnarray}
\langle \frac{1}{z} \rangle
=
\frac{\int dz\frac{1} {z^2} D_{p/c} (z) z^{4+1/2} \frac{1}{z}}
{\int dz\frac{1} {z^2} D_{p/c} (z) z^{4+1/2} }.
\end{eqnarray}
We can obtain the magnitude of $\langle1/ z\rangle $ by using the BKK
fragmentation functions from Ref.\ \cite{BKK} for a parton to fragment into a
pion for $Q_0^2=2$ GeV$^2$,
\begin{eqnarray}
D_{\pi/q}(z) &=& 0.551z^{-1} (1-z)^{1.2},\nonumber\\
D_{\pi/g}(z) &=& 3.77 (1-z)^2.\nonumber
\end{eqnarray}
We find
\begin{eqnarray}
\bar c_T= p_T\langle\frac{1}{ z} \rangle= \begin{cases}
2.2 ~p_T, & {\rm for~a~ gluon~parton,} \cr
2.46 ~p_T, & {\rm for~a~ quark~parton} .\cr
\end{cases}
\end{eqnarray}
For our numerical work, we shall use the average value for gluon and
quark partons,
\begin{eqnarray}
\bar c_T&=& p_T\langle\frac{1}{ z} \rangle=  2.33  ~p_T.
\label{eq65} 
\end{eqnarray}
The differential cross section ${d\sigma (AB\to pX) }/{dyd\bb p_T }$
of Eq.\ (\ref{eqA61}) for the hard scattering of hadrons $A$ and $B$
after fragmenting (and showering) to hadron $p$ can be approximated by
\begin{eqnarray}
&&\frac {d\sigma (AB\to pX) }{dy d\bb p_T}
\nonumber\\
&& \hspace*{0.7cm}\propto \frac{
\alpha_s^2(\bar c_T)(1-x_{a0}(\bar c_T))^{g_a}(1-x_{b0}(\bar c_T))^{g_a} 
}{p_T^{4+1/2}},~~~~~~~~   
\label{eq60a}
\end{eqnarray}
where $\bar c_T$ is given by Eq.\ (\ref{eq65}).

\section {Parton Showering and the Power Index $n$}

The results of the last section indicate  that with a fragmentation
fraction $z$ that is independent of the fragmentation parton momentum
$c_T$ in the showering process, the power law and the power index are
unchanged, and the power index $n+1/2$ for the produced hadrons should
be approximately 4.5 as given by Eq.\ (\ref{AB}) or Eq.\ (\ref{eq60a}).  On
the other hand, the transverse spectra of produced hadrons in
high-energy $pp$ collisions at the LHC gives a power index $n$$\sim$7
\cite{CMS11,CMS12a,ATLAS,ALICE,Won12,Arl10}.  Theoretically, the
PYTHIA program with additional parton showering and radiations, can
describe quite well the transverse momentum distributions of produced
hadrons in $pp$ collisions at LHC energies \cite{CMS11}, which are
associated with a power index $n$$\sim$7 \cite{Won12}.  The difference
between the power index of $n$$\sim$4-5 from the transverse
differential cross sections
of
hadron and photon jets and $n$$\sim$7 from the transverse spectra
of hadrons is likely to arise from the subsequent showering of the
parton jets to hadron fragments of lower transverse momenta.

It should be realized that the showering mechanism presented in the
last section may not contain sufficient degrees of freedom to describe
properly the QCD showering process.  In addition to the kinematic
decrease of the magnitude of the transverse momentum as governed by
Eq.\ (\ref{eq56}),
\begin{eqnarray}
\frac{p_T}{c_T} =   \zeta ^\lambda,
\label{60}
\end{eqnarray}
the showering is governed by an additional criterion on the
virtuality, which measures the degree of the off-the-mass-shell
property of the parton.  There are three different parton showering
schemes: the PHYTHIA \cite{PYT}, the HERWIG \cite{HER}, and the
ARIADNE \cite{ARI}.  The general picture is that the initial parton
with a large initial virtuality $Q$ decreases its virtuality by
showering until a limit of $Q_0$ is reached. Each of the three schemes
uses a different relation between the virtuality and the attributes of
the showering parton, and each with a different evolution variable and
a different virtuality limit.  Their kinematical schemes, the
treatments of soft gluon interference, and the 
hadronization schemes are also different.
  
We can abstract from these different parton showering schemes to
infer that there is approximately a one-to-one mapping of the initial
virtuality $Q$ with the transverse momentum $c_T$ of the evolving
parton as showering proceeds.  The initial virtuality $Q$ scales with,
and maps into, the initial transverse $c_T$ of the showering parton,
and the cutoff virtuality $Q_0$ scales with, and maps into, a
transverse momentum $p_{T0}$ of the parton. In each successive
generation of the showering, the virtuality decreases by a virtuality
fraction which corresponds, in terms of the corresponding mapped
parton transverse momentum, to a decrease by a transverse momentum
fraction $\tilde \zeta$.  The showering will end in $\lambda$
generations such that
\begin{eqnarray}
\frac{p_{T0}}{c_T} = a\tilde \zeta^\lambda,
\end{eqnarray}
where $a$ is a constant relating the scales of virtuality and
transverse momentum.  Thus, the showering process depends on the
magnitude of $c_T$ and the limiting virtuality $Q_0$, which
corresponds to a parton momentum $p_{T0}$.  The greater the value of
$c_T$, the greater the number of generations $\lambda$.  We can infer
an approximate relation between $c_T$ and the number of generations
$\lambda$,
\begin{eqnarray}
\lambda={\ln \frac{p_{T0}}{ac_T} }\biggr / {\ln \tilde \zeta}.
\end{eqnarray}
On the other hand, kinematically, the showering processes degrade the
transverse momentum of the parton $c_T$ to that of the hadron $p_T$ as
given by Eq.\ (\ref{60}), depending on the number of generations
$\lambda$.  The magnitude of the hadron transverse momentum $p_T$ is
related (on the average) to the parton transverse momentum $c_T$ by
\begin{eqnarray}
\frac{p_T}{c_T} = \zeta^{\lambda}= \zeta ^{{\ln \frac{p_{T0}}{ac_T} } /{\ln \tilde \zeta}}.
\end{eqnarray}
We can solve the above equation for $p_T$ as a function of $c_T$, 
\begin{eqnarray}
\frac{p_T}{p_{T0}} 
&=&\left ( \frac{c_T}{p_{T0}} \right )
 ^{1- \mu} a^{-\mu},
\end{eqnarray}
and alternatively for $c_T$ as a function of $p_T$, 
\begin{eqnarray}
\frac{c_T}{p_{T0}} 
&=&\left ( \frac{p_T}{p_{T0}} \right )
 ^{1/(1- \mu)}a^{\frac{\mu}{1-\mu}},
\label{65}
\end{eqnarray}
where 
\begin{eqnarray}
\mu=\ln \zeta /{\ln \tilde \zeta} >0,
\end{eqnarray}
and $\mu$ is a parameter that can be searched to fit the data.  As a
result of the virtuality ordering and virtuality cut-off, the hadron
fragment transverse momentum $p_T$ is related to the parton momentum
$c_T$ by an exponent $1-\mu$.

After the fragmentation and showering of the parton $c$ to hadron $p$,
the hard-scattering cross section for the scattering in terms of
hadron momentum $p_T$ becomes
\begin{eqnarray}
&&
 \frac{d^3\sigma( AB \to p X) }{dy d{\bb p}_T}
= \frac{d^3\sigma( AB \to c X) }{dy d{\bb c}_T}
\frac{d{\bb c}_T}{d{\bb p}_T}
\nonumber\\
&& \hspace*{1.0cm}
\propto \frac{\alpha_s^2(\bar c_T)(1\!-\!x_{a0}(\bar c_T))^{g_a}(1\!-\!x_{b0}(\bar c_T))^{g_a} }
{c_T^{4+1/2}}\frac{d{\bb c}_T}{d{\bb p}_T}.~~~~
\end{eqnarray}
From the relation between the parent parton moment $c_T$ and the
leading hadron $p_T$ in Eq.\ (\ref{65}), we get
\begin{eqnarray}
\frac{d{\bb c}_T}{d{\bb p}_T}
&=&
{\frac{1}{1- \mu}}\left ( \frac{p_T}{p_{T0}} \right )
 ^{\frac{2\mu}{1- \mu}}a^{\frac{2\mu}{1-\mu}}.
\label{77}
\end{eqnarray}
Therefore under the fragmentation from $c$ to $p$, the hard-scattering
cross section for $AB \to p X$ becomes
\begin{eqnarray}
 \frac{d^3\sigma( AB \to p X) }{dy d{\bb p}_T}
\!\propto \!\frac{\alpha_s^2(\bar c_T)(1\!-\!x_{a0}(\bar c_T))^{g_a}(1\!-\!x_{b0}(\bar c_T))^{g_a} }
{p_T^{n'}},~~~
\nonumber\\
\label{fnl}
\end{eqnarray}
where
\begin{eqnarray}
n'&=&\frac{n-2\mu}{1-\mu},~~{\rm with~~} n=4+\frac{1}{2}.
\end{eqnarray}
Thus, from Eqs.\ (\ref{65})-(\ref{77}), the parton showering process
with limiting virtuality may modify the power law index in the
transverse differential cross section from $n$ to $n'$.  The parameter
$\mu$ is related to $n$ and $n'$ by
\begin{eqnarray}
\mu=\frac{n'-n}{n'-2}.
\end{eqnarray}

\section{Phenomenological Modifications of the Hard-Scattering cross section}

In the last section we give qualitative arguments to show that the
power index may be modified from $n$ to $n'$ by the process of
showering. A quantitative evaluation of the changes in the power index
from fundamental QCD principles is difficult, because the showering
and the subsequent hadronization processes are complicated and contain
unknown nonperturbative elements.  It suffices to verify that there
is indeed a systematic change of the power index from partons (or their
equivalent representative jets) to hadrons, by finding the empirical
values of power index $n$ for hadron production.  For such a purpose, we shall
modify the differential cross section $d^3\sigma( AB \to p X) /dy
d{\bb p}_T$ in (\ref{fnl}), for an incident parton $a$ scattering into
$c$ after a relativistic hard scattering, showering, and hadronization
to be
\begin{eqnarray}
 \frac{d^3\sigma( AB \to p X) }{dy d{\bb p}_T}
\propto \frac{\alpha_s^2(\bar c_T)(1\!-\!x_{a0}(\bar c_T))^{g_a}(1\!-\!x_{b0}(\bar c_T))^{g_a} }
{[1+m_{T}/m_{T0}]^{n}}, 
\nonumber\\
\label{eq76},~~~
\end{eqnarray}
where $m_T$ is the transverse mass $\sqrt{m^2+p_T^2}$ of the detected
hadron $p$, and $m$ is the hadron mass taken to be the pion mass.  The
transverse mass $m_{T0}$ has been introduced both to regulate the
behavior of the cross section in the region of small $p_T$ and to
represent the average transverse mass of the detected hadron in the
hard-scattering process.
\begin{figure} [h]
\includegraphics[scale=0.45]{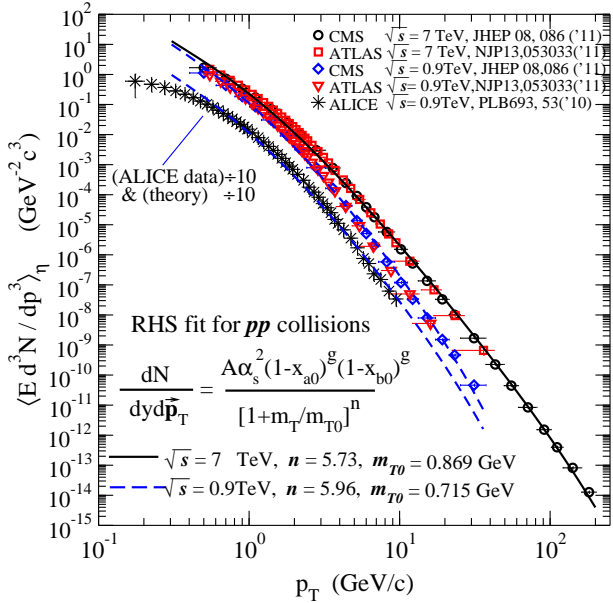}
\caption{(Color online) Comparison of the experimental transverse
  momentum distribution $\langle E_p d^3N/dp^3\rangle_{\eta}$ of
  hadrons in $pp$ collisions with the relativistic hard-scattering
  model Eq.\ (\ref{43}), assuming a linear $m_T$ dependence of the
  regulating function.}
\end{figure}
Experiments measure the differential yield in non-single-diffractive
events, which is related to the differential cross section by
\begin{eqnarray}
E_p \frac{d^3 N(AB \to pX)}{ d p^3}
=E_p \frac{d^3 \sigma (AB \to pX)}{ \sigma_{\rm NSD}d p^3},
\end{eqnarray}
where $\sigma_{\rm NSD}$ is the non-single-diffractive cross-section.
We also need to transcribe the invariant cross section in terms of
$d\sigma/d\eta d\bb p_T$.  We have then the produced particle
distribution
\begin{eqnarray}
&&\frac{d^3 N(AB \to pX)}{ d\eta  d\bb p_T}
= \sqrt{1-\frac{m^2}{m_T^2\cosh^2 y}} \nonumber\\
&& \hspace*{1.0cm}
\times
A 
\frac{ \alpha_s^2({\bar c_T})(1-x_{a0}({\bar c_T}))^{g_a}(1-x_{b0}({ \bar c_T}))^{g_b}}
{[1+m_{T}/m_{T0}]^{n}},~~~~~
\label{43}
\end{eqnarray}
where $A$ is a constant fitting parameter.  We shall use the above
formula Eq.\ (\ref{43}) to search for the power index $n$ for hadron
production by fitting the hadron transverse momentum distributions in
$pp$ collisions at LHC from the CMS \cite{CMS12a}, ATLAS \cite{ATLAS},
and ALICE collaborations \cite{ALICE}, within the experimental
pseudorapidity windows.  We shall again take $g_a=g_b=6$ \cite{Duk84}.
In Fig.\ 6, we compare the fits to the experimental hadron transverse
spectra.  We find that for $pp$ collisions at $\sqrt{s}$=7 TeV, the parameters are 
 $n=5.73$, $m_{T0}=0.869$ GeV, and $A=194$ GeV$^{-2}c^3$, and for $pp$
collisions at $\sqrt{s}$=0.9 TeV,  the parameters are  $n=5.96$,
and $m_{T0}=0.715$ GeV, $A=236$ GeV$^{-2}c^3$.

Note that if we introduce
\begin{eqnarray}
q=1+\frac{1}{n} {\rm ~~~and~~} T=\frac{m_{T0}}{q-1},
\label{eq50}
\end{eqnarray}
then we get 
\begin{eqnarray}
&&\frac{d^3 N(AB \to pX)}{ d\eta  d\bb p _T}
= \sqrt{1-\frac{m^2}{m_T^2\cosh^2 y}}
\nonumber\\
&&\hspace*{1.5cm}
\times
 A 
{ \alpha_s^2({\bar c_T}) (1-x_{a0}({\bar c_T}))^{g_a}(1-x_{b0}({\bar c_T}))^{g_b}}~~~~
\nonumber\\
&&\hspace*{1.5cm}\times
{\left [1-(1-q)\frac{m_{T}}{T}\right ]}^{\frac{1}{1-q} },
\label{50}
\end{eqnarray}
which is in the form of the Tsallis distribution of Eq. (1) (now with
a clear meaning of the ``nonextensivity parameter" $q$ and the
``temperature"$T$ as given in Eq.\ (\ref{eq50})). The difference is
the additional $p_T$ dependencies of $\alpha_s^2(\bar c_T)$,  $x_{a0}(\bar c_T)$,
$x_{b0}(\bar c_T)$ as well as the square-root prefactor.  What needs to be
stressed is that the real active number of degrees of freedom remains
quite small, similar to Eq. (1).

Equation (\ref{eq76}) is not the only way we can parametrize the
hard-scattering results.  The gluon exchange propagator in the Feynman
diagrams of Figs. 1 and 2 and Eqs.\ (\ref{eq29}) and (\ref{eq33})
involve the quantities $q_{Ti}^2$. We can alternatively modify the
basic differential cross section $d^3\sigma( AB \to p X) /dy d{\bb
  p}_T$ for the scattering of $a$ to $p$ in the quadratic $m_T^2$
form,
\begin{eqnarray}
&& \frac{d^3\sigma( AB \to p X) }{dy d{\bb p}_T}
\nonumber\\
&& \hspace*{1.3cm}
\propto \frac{ \alpha_s^2({\bar c_T})(1\!-\!x_{a0}(\bar c_T))^{g_a}(1\!-\!x_{b0}(\bar c_T))^{g_a} }
{[1+m_{T}^2/m_{T0}^2]^{n/2}}.~~~~~~~
\end{eqnarray}
With such an effective representation of the basic $a\to p$
scattering, Eq. (\ref{fnl}) is altered to become
\begin{eqnarray}
&&\!\!\!\!\!\!\!\!\!\!\!\frac{d^3 N(AB \to pX)}{ d\eta  d\bb p _T}
= \sqrt{1-\frac{m^2}{m_T^2\cosh^2 y}} 
\nonumber\\
&& ~~~~~~\times
A
\frac{\alpha_s^2({\bar c_T}) (1-x_{a0}({\bar c_T}))^{g_a}(1-x_{b0}({\bar c_T}))^{g_b}}
{[1+m_{T}^2/m_{T0}^{2}]^{n/2}}.
\label{49}
\end{eqnarray}

\begin{figure} [h]
\includegraphics[scale=0.45]{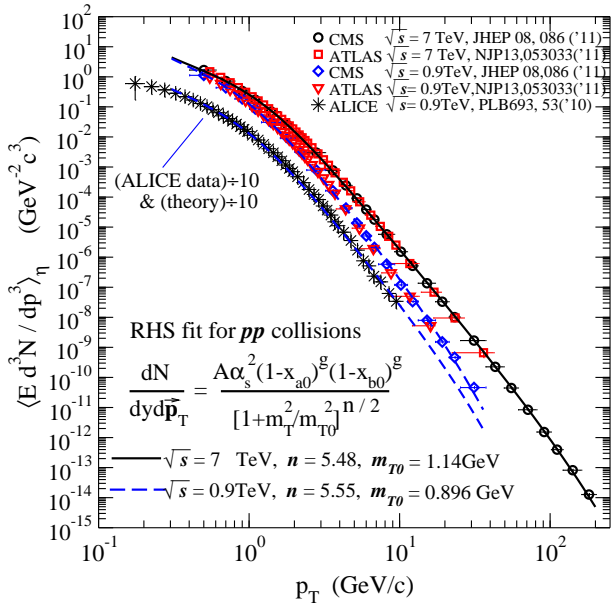}
\caption{(Color online) Comparison of the experimental hadron
  transverse momentum distribution $\langle E_p
  d^3N/dp^3\rangle_{\eta}$ of hadrons in $pp$ collisions with the
  relativistic hard-scattering model Eq.\ (\ref{43}), assuming a
  quadratic $m_T$ dependence of the regulating function.}
\end{figure}

We use the above equation with the quadratic $m_T^2$ dependence in the
transverse distribution to search the power index $n$ by  fitting  the experimental hadron transverse
momentum distribution $\langle E_p d^3N/dp^3\rangle_{\eta}$ in $pp$
collisions from the CMS\cite{CMS12a}, ATLAS\cite{ATLAS}, and ALICE
collaborations\cite{ALICE}.  The data for $p_T \agt 0.5$ GeV/c agree
with the theoretical fits as shown in Fig.\ 8.  The parameters
for $pp$ collisions at $\sqrt{s}=7$ TeV are 
$n$=5.83, $m_{T0}=0.856$ GeV, and $A$=3.58 GeV$^{-2}c^3$, and for $pp$ collisions at
$\sqrt{s}=0.9$ TeV, the parameters are 
$n$=5.97, $m_{T0}=0.685$ GeV, and $A$=4.58 GeV$^{-2}c^3$.    We give the
fitting parameters that describe the $p_T$contributions from spectra
at the two different energies in Table II.

\vspace*{0.3cm} 
\begin{table}[h]
\caption { Fitting parameters  $n$, $m_{T0}$, and $A$ for the
  transverse momentum distribution of hadrons in $pp$ collisions.  
  }
\vspace*{0.3cm}
\begin{tabular}{|c|c|c|c|c|}
\cline{2-5}
   \multicolumn{1}{c}       {}        &  \multicolumn{2}{|c|} {Linear $m_T$}   & \multicolumn{2}{c|} {Quadratic $m_T^2$}  \\
   \multicolumn{1}{c}       {}        &  \multicolumn{2}{|c|} { Eq.\ (\ref{43})}   & \multicolumn{2}{c|} {Eq.\ (\ref{49}) }  \\
\cline{2-5}
       \multicolumn{1}{c|} {}               & $\sqrt{s}$=7TeV&  $\sqrt{s}$=0.9TeV  & $\sqrt{s}$=7TeV  & $\sqrt{s}$=0.9TeV  
 \\ \hline
$n$    & 5.73 &  5.96 &   5.48&  5.55   \\ \hline
$m_{T0}$ (GeV)  & 0.869 & 0.715 &   1.14  &  0.896   \\ \hline
$A$(GeV$^{-2}c^3$)& 194 &  236    &  12.8&  13.8   \\ \hline
\end{tabular}
\end{table}

Comparing the results from the two different ways of expressing the
power-law behaviors, we find that the agreements of the data with the
theoretical curves are nearly the same above $p_T$$\agt $3 GeV/c, but
the theoretical results for the linear case with the $m_T$ dependence of
Eq.\ (\ref{43}) are less than the experimental ALICE data for $p_T$$\sim$2 GeV/c but greater than the experimental data for $p_T$$\alt $0.5 GeV/c.  On the other hand,
the quadratic $m_T^2$ expression of Eq.\ (\ref{49}), that is 
a more natural from field theory point of view involving
gluon propagators,  leads to a better agreement in the lower $p_T$ region.

For $pp$ collisions at the LHC, the above comparisons indicate that the
power index extracted from hadron spectra has the value of $n$$\sim$6.
The power index is systematically larger than the power index of
$n$$\sim$4-5 extracted from jet transverse differential cross sections.
Considering the difference of a jet and  hadrons, we can infer that the process of fragmentation and
showering increases the value of the power index $n$ of the
transverse spectra.

It should be noted that the hard-scattering model results in the
low-$p_T$ region will be slightly modified with the introduction of
the intrinsic $p_T$ of the partons \cite{Won98}.  There will also be
modifications due to the recombination of partons \cite{Hwa03}.
Nevertheless, the extrapolation of the hard-scattering results to the
low-$p_T$ region as obtained here indicates indeed that the
hard-scattering process can contribute substantially to the production
of particles at the low-$p_T$ region\footnote{Note that, for example,
  for $q \neq 1$ the normalization of the rapidity distribution given
  by Eq. (\ref{50}) depends on $q$.} as has been suggested by Trainor
and collaborators \cite{Tra08}.

\section{Discussions and Conclusions}

We have been stimulated by the good agreement of the Tsallis
distribution with the transverse momentum distribution of produced
hadrons over a large range of the transverse memorandum in $pp$
collisions at LHC energies.  The simplicity of the Tsallis
distributions raises questions on the physical meaning of the few degrees
of freedom entering into the Tsallis distribution.

As the magnitude of the transverse momentum in this high-$p_T$ region
is much greater than the mean transverse momentum, concepts such as
statistical mechanics that depend on thermodynamical equilibrium or
quasiequilibrium may be subject to question.  The asymmetry between
the transverse and the longitudinal degrees of freedom also poses
additional difficulties in a statistical explanation of the full
three-dimensional momentum distribution in this high-$p_T$ region.

We therefore attempt to understand the results of simple Tsallis fit
of the transverse momentum distribution in $pp$ collisions within the
relativistic hard-scattering model.  The relativistic hard-scattering
model however predicts that the differential cross section for the
production of high-$p_T$ particles should vary as $1/p_T^n$ with $n=4$
if the basic process consists of elementary parton-parton $2\to 2$
processes.  The Tsallis fit to the LHC data gives a power index for
hadrons of $n$$\sim$7 that is substantially greater.

Our reexamination of the relativistic hard-scattering model reveals
that for minimum biased events without a centrality selection, the
differential cross section at high $p_T$ is dominated by the
contribution from a single parton-parton collision with the $\alpha_s^2/c_T^4$
behavior.  The multiple scattering process leads to contributions of
higher power indices  that will not modify
significantly the $\alpha_s^2/c_T^4$ behavior at high $p_T$. The power index $n$ should be
approximately 4+1/2 where the additional power of 1/2 arises from the
integration of the structure function.  Indeed, comparison with the
experimental power indices in the transverse differential cross
sections for jet production supports the approximate validity of a
basic $\alpha_s^2/c_T^4$ behavior for parton-parton collisions in relativistic
hard-scattering processes.

As a hadron jet or a photon jet corresponds to the state of a parton
after a parton-parton collision but before the final-state showering,
we now understand that the systematic difference between the power
index of $n$$\sim$4-5 for jets
\cite{Abe93,Abb01,Aco02,Abb00,Aba01,Arl10,Alice13}, and $n$$\sim$6-7
for hadrons \cite{Won12} may be attributed to the subsequent showering
and hadronization of the parton jet to hadron fragments of lower
transverse momenta.  Another part of the increase of the power index
arises from the $p_T$ dependence of the structure function factor $(1-x_{a0})^g(1-x_{b0})^g$ and the running coupling constant.

While we examine here the contributions of the hard processes, there
can also be contributions of the produced particles from soft
processes in the low-$p_T$ region.  These contributions relative to
those from hard processes will certainly diminish as the collision
energy increases.  It is therefore entirely possible that the
borderline between soft and hard processes moves to the lower $p_T$
region as the collision energy increases.  How the borderline between
the two processes can be determined will require much more future
work.

Many relevant questions on the borderline between the high-$p_T$ and
the low-$p_T$ regions will need to be settled in the future.  First,
it is expected that hard-scattering processes will be accompanied by
collisional correlations different from those from soft processes.  A
careful analysis of the two-particle correlations in the low-$p_T$
region may provide a way of separating out the soft process
contributions from the hard-scattering collisional contributions in
the low-$p_T$ region \cite{Tra08}.  Second, while we apply the
relativistic hard-scattering model to the low-$p_T$ region of $p_T\alt
2$ GeV/c, the approximations we have used may not have its range of
validity down to such regions.  The establishment of the low-$p_T$
limit of validity of the relativistic hard-scattering model will be
both an experimental and theoretical question.  Processes such as
parton intrinsic transverse momentum \cite{Won98} and parton
recombination \cite{Hwa03} will add complexity to the transverse
momentum distribution in the low-$p_T$ region.  Third, the
separation of the soft process contribution and the knowledge of the
borderline between the soft processes and the hard processes may also
provide information whether the basic collision law should be
represented by a linear form of $m_T$ in Eq. (\ref{43}) or a quadratic
form of $m_T^2$ in Eq. (\ref{49}). 

The low-$p_T$ region is conventionally associated with soft
nonperturbative processes and the high-$p_T$ region with perturbative
hard-scattering processes.  A very different two-component Model (TCM)
scheme for partitioning the soft and hard components has been proposed
\cite{Ada06,Tra08}.  Measurements of the STAR collaboration
\cite{Ada06} on the transverse distribution $d^3N/d\eta dp_T^2$ around
$\eta$$\sim$0, as a function of the event multiplicity classes, reveal
that the distribution $d^3N/d\eta dp_T^2$ can be approximately written
as the sum of a term linear in multiplicity, $n_{\rm ch} S_0(p_T)$,
and a term quadratic in multiplicity, $n_{\rm ch}^2 H_0(p_T)$
\cite{Ada06}.  Under the hypothesis that the multiplicity of hard
collisions $n_{ h}$ is proportional to $n_{\rm ch}^2$ while the
multiplicity of soft collisions $n_{s}$ is linear in $n_{\rm ch}$, the
$S_0(p_T)$ contribution, parametrized in the Levy form or the
equivalent Tsallis form as a function of $p_T$,
$S_0(p_T)$$\sim$$1/[1+(m_T-m_0)/nT]^n$, is identified in the TCM
scheme as the TCM ``soft" component, and the $H_0(p_T)$ contribution,
parametrized as a Gaussian in shifted $y_T=\ln [( m_T + p_T)/m]$, is
identified as the TCM ``hard" component \cite{Ada06,Tra08}.  As a
result of such a partition, the TCM soft component remains significant
even at very high $p_T$ and contains a power law $1/p_T^n$ behavior,
which however occurs only in the conventional hard component of
relativistic hard-scattering model.  On the other hand, the TCM hard
component is a Gaussian distribution in shifted $y_T$ centered at
$p_T\sim1.4$ GeV and it does not have the power-law behavior of
relativistic hard scattering model at high $p_T$. The TCM partitions
are in variance with those in our physical, and conventional
partitions.  Furthermore, from physical arguments, one expects that
the multiplicity of relativistic hard-scattering collisions $n_h$ need
not be related to the square of the multiplicity of soft collisions
$n_s^2$, and the soft and hard processes contribute in different
regions of $p_T$.  Constraining $n_h$ to be proportional to $n_s^2$ in
the partition may lead to a distortion of the spectrum of TCM
components.  As there are many different ways of partitioning the
spectrum, the theoretical, physical, and mathematical basis for the
two-component model partition in the form as presented as soft and
hard in \cite{Ada06,Tra08} may need to be further investigated. 

Returning to the Tsallis distribution which motivates the present
investigation, we can conclude that the successes of representing the
transverse spectra at high-$p_T$ by a Tsallis distribution arise from
(i) the simple power-law behavior of the parton-parton scattering
cross section, $\alpha_s^2/c_T^4$, with a power index of $4$, and (ii)
the few number of the degrees of freedom in the hard-scattering model.
The power index of 4 has been found experimentally to be approximately
valid by examining the differential cross sections of hadron jets and
photon jets.  It has also been found theoretically to be approximately
valid by examining the multiple scattering process.  The power index
is not significantly modified by the multiple scattering process in
minimum biased measurements.  The $\alpha_s^2/p_T^4$ power law lays
the foundation for Tsallis/Hegedorn-type transverse momentum
distributions, and the few degrees of freedom in the Tsallis
distribution is a reflection the few degrees of freedom in the
underlying hard-scattering model.  There are additional $p_T$
dependence due to the parton structure function, the running coupling
constant, and the parton momentum integration, which lead to a
slightly larger power index.  Furthermore, in going from the parton
measurements in terms of jets to hadron measurements in terms of
fragmented hadron products, there are additional showering and
fragmentation processes which give rise to a greater value of the
power index.  The Tsallis distribution is flexible enough to adjust
the power index to accommodate the different and changing environment,
yielding a nonstatistical description of the distribution.

Because of its nonstatistical nature, the parameters in a Tsallis
distribution can only be supplied and suggested from nonstatistical
means, such as the QCD basic parton-parton scattering power index and
the QCD multiple scattering shadowing effects.  It also is limited in
its application to the transverse degree of freedom, as there is no
way to generalize the Tsallis parameters across the three-dimensional
space from transverse to longitudinal coordinates.  For a more
fundamental description, it is necessary to turn to the basic parton
model for answers.  For example, the relativistic hard scattering model can
be applied to collision to other longitudinal regions of
pseudorapidities where in the forward rapidity region, the additional
mechanism of direct fragmentation \cite{Won80} should also be
included.  The underlying relativistic hard-scattering model has a
greater range of applications and a stronger theoretical foundation.

\vspace*{0.3cm} \centerline{\bf Acknowledgment}

\vspace*{0.3cm} The authors would like to thank Profs.
R. Blankenbecler, Vince Cianciolo, R. Hwa, Jiangyong Jia,
D. Silvermyr, T. Trainor, and Z.~W\l odarczyk for helpful discussions
and communications.  The research was supported in part by the
Division of Nuclear Physics, U.S. Department of Energy (C.Y.W) and by
the Ministry of Science and Higher Education under Contract
DPN/N97/CERN/2009 (G.W.).

\end{document}